\documentclass[10pt,twocolumn]{article}

\usepackage{graphicx}
\usepackage[hyphens]{url}
\usepackage{amsmath,amsfonts,amssymb,amsthm,bm,mathrsfs}

\usepackage[letterpaper, total={7.2in, 9.2in}, columnsep=0.25in]{geometry}

\numberwithin{equation}{section}

\usepackage{subcaption}
\usepackage{easybmat}
\usepackage[colorinlistoftodos,textsize=footnotesize]{todonotes}

\newtheorem{theorem}{Theorem}
\newtheorem{lemma}[theorem]{Lemma}
\newtheorem{proposition}[theorem]{Proposition}

\def\birth{b}

\def\IR{\mathbb{R}}

\def\D{\mathcal{D}}
\def\M{\mathcal{M}}
\def\R{\mathcal{R}}

\def\ba{\bm{a}}
\def\bb{\bm{b}}

\def\bu{\bm{u}}

\def\bI{\bm{I}}
\def\bN{\bm{N}}
\def\bR{\bm{R}}
\def\bS{\bm{S}}
\def\bV{\bm{V}}
\def\bX{\bm{X}}

\def\bZero{\bm{0}}

\def\diag{\textrm{diag}}

\usepackage{tikz}
\usetikzlibrary{shapes,arrows}
\usetikzlibrary{positioning}
\tikzstyle{cloud} = [draw, ellipse,fill=red!20, node distance=0.87cm,
minimum height=2em]
\tikzstyle{line} = [draw, -latex']
\usetikzlibrary{shapes.symbols,shapes.callouts,patterns}
\usetikzlibrary{calc}

\begin{document}

\title{A resident-traveller model for global aspects of vaccination}
\author{ {\sc Julien Arino}\\[2pt]
Department of Mathematics, University of Manitoba, \\
Winnipeg, Manitoba R3T 2N2, Canada.\\[6pt]
{\sc P. van den Driessche}\\[2pt]
Department of Mathematics and Statistics, University of Victoria, \\
Victoria, British Columbia V8W 2Y2, Canada.\\[6pt]}
\pagestyle{headings}
\maketitle

\begin{abstract}
A model for the spread of a vaccine-preventable infectious disease is set in a resident-traveller metapopulation context to investigate global aspects of vaccination. 
Of particular interest here is the effect of varying vaccine coverage in some patches on the spread of disease in the whole system.
Analytically and numerically, the resident-traveller framework is compared with a classic metapopulation model where individuals become residents of the patch they move to.
A numerical example inspired by the 2003--2004 poliomyelitis resurgence in western and central Africa is then considered, illustrating how heterogeneous vaccine coverage can sustain widespread regional transmission that classic metapopulation models predict to be much smaller.

\end{abstract}

\section*{Introduction}
\label{sec:intro}
The increase and acceleration of travel has dramatically eased the global dissemination of infectious pathogens. This was made clear when SARS spread across the globe in 2003 \cite{ChristianPoutanenLouftyEtAl2004,RuanWangLevin2006}, with the propagation of pandemic H1N1 influenza in 2009 \cite{KhanArinoHuEtAl2009} and, most recently, with the global spread of SARS-CoV-2 and its variants \cite{Arino2022}.
However, aside from the rapid spread of emerging or reemerging infectious diseases, globalization of human movement poses other problems. 
One of them is that it is becoming harder to devise public health policies within the restricted span of individual countries or regions and, as a consequence, to consider policies that do not take into account the practices of other health jurisdictions.
While public policies are typically formulated and conducted within a country or local public health jurisdiction, the movement of individuals between countries or jurisdictions implies that issues or deficiencies in the public health system of one jurisdiction can have consequences in others.

For instance, the 2009 influenza pandemic was under way when the yearly pilgrimage to Mecca took place. 
The gathering of pilgrims from many countries in these circumstances led to much elevated risk of importation of the disease to countries that were at that time almost unaffected by it \cite{KhanMemishChabbraEtAl2010}, although this particular pandemic turned out to be generally benign enough that this did not cause major issues.
The use of antivirals against influenza during the 2009 pandemic varied greatly from country to country \cite{BereraZambon2013}. 
While some countries were attempting to curtail a potential surge of resistance against antivirals, others were not, so that it is not clear what consequences could have been expected of the policies of the former had widespread resistance indeed emerged. 
Tuberculosis is another disease where the difference between health systems plays a role. 
In Canada, for instance, the four countries providing the most new permanent residents in recent years are China, India, Pakistan and the Philippines. 
These four countries have high tuberculosis burden and not surprisingly, the rate of tuberculosis among new migrants to Canada is much higher than it is for the Canadian-born population \cite{LangloisklassenWooldrageManfredaEtAl2011}.
Similarly, in the United States, because latent tuberculosis prevalence and drug-resistance profiles among arriving migrants reflect conditions in their countries of origin \cite{Khan2002}, decision-analysis studies have shown that it would be cost-saving for the United States to invest directly in expanding tuberculosis control programmes in high-incidence source countries \cite{Schwartzman2005}.
However, this type of recommendation is rarely taken into account by public health authorities.

Another aspect of public health that is subject to spatial and temporal effects is vaccination.
The motivation for the present work rests in the fact that there is evidence that, in 2003--2004, the interruption of poliomyelitis vaccination in the northern Nigerian state of Kano \cite{ClementsGreenoughShull2006,Ghinai2013,Jegede2007} led over the course of several months to an increase of the number of cases of polio in neighbouring regions \cite{PallanschSandhu2006}.
Vaccination then resumed \cite{Fleck2004a}, but problems persisted long after this event \cite{KaufmannFeldbaum2009}.
Thus, eradication efforts, when they are conducted, should attack problems at a global scale \cite{KlepacMetcalfMcLeanHampson2013}.
At a more local scale, there have also been several outbreaks of childhood and other vaccine-preventable diseases in countries where vaccine coverage is generally high \cite{KnolUrbanusE.M.MollemaEtAl2013}. 
This is a worrying trend: because of misinformation about vaccines \cite{DownsBruinFischhoff2008} or religious convictions \cite{Grabenstein2013}, the number of individuals who are not vaccinated is on the rise. 
This phenomenon happens at the local scale and can even be driven by varying physicians' attitudes towards vaccination \cite{WeigelWeitmannRautmannSchmidtEtAl2014}.
Vaccination poses interesting ethical issues \cite{IsaacsKilhamLeaskTobin2009} as it lies at the intersection of the public and private spheres. 
Debates concerning vaccination are likely to persist \cite{BedfordElliman2000}.
The reasons that lead to refusal of vaccination are diverse \cite{Blume2006}, but whatever the reasons, the increased prevalence of the Internet as a place for sharing often incorrect information \cite{Kata2010,Kata2012} and the persistence of anti-vaccination movements means that it is of considerable importance to understand the effect of varying vaccine coverage levels.

With these examples in mind, in order to illustrate the effect of local policies on global health, the effect of vaccination is investigated in a metapopulation model that describes travel of individuals between patches or regions. 
The purpose here is not to formulate full polio or measles vaccination models. 
Such models would indeed need to include other characteristics of the virus and vaccines that for simplicity are not considered here. 
Previous work has considered the effect of spatial aspects of vaccination for specific diseases.
In particular, \cite{KlepacMetcalfMcLeanHampson2013} focuses on disease elimination when infections are few, e.g., polio, and \cite{MetcalfCohenLesslerMcAnerneyEtAl2013} uses data on rubella in South Africa to consider the outcome of heterogeneous vaccination.

The present work uses a model framework of travel between patches introduced in \cite{SattenspielDietz1995}, used in \cite{SattenspielHerring2003} and others and studied analytically in \cite{ArinoVdD2003b,ArinoVdD2003a,IggidrSalletTsanou2012}. 
This type of model is sometimes called a \emph{mover-stayer} or \emph{resident-traveller} model (the latter name being used here) because it keeps track of the place of residence of individuals.
As in further work of the authors with this model \cite{Arino2009,ArinoVdD2006a}, the original model of \cite{SattenspielDietz1995} is extended to allow individuals to travel between remote patches.
It is established that the behaviour of the resulting travel model, \emph{i.e.}, the model describing the dynamics of the total population in each subpopulation, is similar to that in \cite{ArinoVdD2003b,ArinoVdD2003a}, namely that with more general travel, each patch population has an asymptotically stable positive equilibrium and the resident population in each patch is asymptotically constant.
A model for vaccination in different patches is then formulated and mathematically and computationally analysed.

This work also aims to compare the resident-traveller formulation with classic metapopulation models. 
In classic models, individuals who travel immediately adopt the parameters and status of their destination. 
This simplification is problematic when modelling jurisdiction-based public health interventions such as vaccination, where individuals are subject to the policies of their residence rather than their temporary location. 
By contrasting the two frameworks, the aim is to show how applying local vaccination rates to visitors in classic models can distort predictions of disease persistence and control.

In the computational investigation of the models, the focus is on two public health questions:
\begin{enumerate}
    \item Can a public health authority that does vaccinate help non-vaccinating neighbouring authorities, or protect its own population?
    \item What happens to the global system when, like in the polio example motivating this work, a jurisdiction ceases to vaccinate?
\end{enumerate}
This is complemented by a numerical thought experiment inspired by the 2003--2004 polio resurgence in western and central Africa.

\section{The mobility model in the absence of disease}
\label{sec:mobility}
\subsection{Formulation of the mobility model}
The model tracks individuals as they move in a multi-digraph comprising $n$ patches (regions) as vertices and links between the patches as (multi) arcs. 
For simplification purposes, it is assumed that an individual who is born in a given patch stays until death a \emph{resident} of this patch. 
This individual may however spend an indefinite amount of time in other patches. 
An individual resident of one patch who is currently in another patch is a \emph{traveller}.
The number of residents of patch $i$ present in patch $j$ at time $t$ is denoted $N_{ij}(t)$; they are referred to as subpopulation $(i,j)$.

The total number of residents of patch $i$ is given by $N^r_i=\sum_{k=1}^n N_{ik}$ and the number of individuals who are physically present in patch $i$ at time $t$, both residents and travellers, is $N_i^p=\sum_{k=1}^n N_{ki}$.
The rates of movement of individuals between patches are written with three ordered subscripts as
\[
m_{\textrm{residence,destination,origin}}.
\]
Thus, individuals from patch $i$ who are visiting patch $j$ travel to patch $k$ at a rate $m_{ikj}\geq 0$. 
The travel structure defines a weighted multi-digraph, with patches as vertices and weighted arcs representing the links between these patches, the weight on the arcs indicating the rates of travel. 
Between any two patches (vertices) there are at most $n$ arcs in each direction. 

Assume that death occurs at the per capita rate $d_{ij}>0$ for residents of patch $i$ currently in patch $j$. 
Birth occurs only in the residence patch, at the rate $\birth_{ii}$.
Dispersion is instantaneous, so death during travel is negligible.

Taking the previous assumptions into account and assuming that population numbers are sufficiently large for a deterministic formulation, ordinary differential equations can be derived for the dynamics of the population.
Firstly, the evolution of the number of residents of patch $i$ present in patch $i$ is given by
\[
\frac{d}{dt}N_{ii}=\birth_{ii}-d_{ii}N_{ii}
+\sum_{\substack{k=1\\ k\neq i}}^n m_{iik}N_{ik}
-\left(\sum_{\substack{k=1\\ k\neq i}}^n m_{iki}\right)N_{ii}.
\]
The evolution of the number of residents of patch $i$ who are present in patch $j\neq i$ is governed by
\[
\frac{d}{dt}N_{ij}=-d_{ij}N_{ij}
+\sum_{\substack{k=1\\ k\neq j}}^n m_{ijk}N_{ik}
-\left(\sum_{\substack{k=1\\ k\neq j}}^n m_{ikj}\right)N_{ij}.
\]
To make the notation more compact, denote
\[
\birth_{ij}=\begin{cases}
\birth_{ii}>0 & \textrm{if }i=j \\
0 & \textrm{if }i\neq j
\end{cases}
\]
the birth constant for subpopulation $(i,j)$ and
\[
m_{ijj}=-\sum_{\substack{k=1\\ k\neq j}}^n m_{ikj}
\]
the rate of movement outbound from patch $j$ for residents of patch $i$.
The general equation for the demographic model is then given for $i,j=1,\ldots,n$ by
\begin{equation}\label{sys:dN}
\frac{d}{dt}N_{ij}=\birth_{ij}-d_{ij}N_{ij}+\sum_{k=1}^n m_{ijk}N_{ik}.
\end{equation}
System \eqref{sys:dN} subject to initial values $N_{ij}(0)\geq 0$ for all $i,j=1,\ldots,n$ and $N_i^r(0)>0$ for all $i=1,\ldots,n$ constitutes the linear nonhomogeneous mobility model in the absence of disease.

\subsection{Behaviour of the total population}
\begin{lemma}\label{lemma:boundsN}
With the given initial conditions, the resident population $N_i^r$ in each patch and the total population are positive and bounded.
\end{lemma}
\begin{proof}
The evolution of the total resident population in patch $i=1,\ldots,n$ takes the form
\begin{equation}\label{eq:dN_nonconstant}
\frac{d}{dt}N_i^r=\frac{d}{dt}\sum_{j=1}^nN_{ij}
=\birth_{ii}-\sum_{j=1}^nd_{ij}N_{ij},
\end{equation}
where the travel terms have cancelled out.
Indeed,
\begin{align*}
\sum_{j=1}^n\left(\sum_{k=1}^{n} m_{ijk}N_{ik}\right) 
&= \sum_{j=1}^n\left(\sum_{\substack{k=1\\ k\neq j}}^n m_{ijk}N_{ik}
-\sum_{\substack{k=1\\ k\neq j}}^n m_{ikj}N_{ij}\right) \\
&= \sum_{j=1}^n\sum_{\substack{k=1\\ k\neq j}}^n m_{ijk}N_{ik}
-\sum_{j=1}^n\sum_{\substack{k=1\\ k\neq j}}^n m_{ikj}N_{ij} \\
&= 0.
\end{align*}
Let $\underline{d}_i=\min_j\{d_{ij}\}$ and $\overline{d}_i=\max_j\{d_{ij}\}$.
Equation \eqref{eq:dN_nonconstant} gives, for all $t\geq 0$,
\begin{multline*}
0<N_i^r(0)e^{-\overline{d}_it}+\frac{\birth_{ii}}{\overline{d}_i}\left(1-e^{-\overline{d}_it}\right)
\leq N_i^r(t) \\
\leq N_i^r(0)e^{-\underline{d}_it}+\frac{\birth_{ii}}{\underline{d}_i}\left(1-e^{-\underline{d}_it}\right).
\end{multline*}
As $N_i^r(0)>0$, the result follows for the total resident population. Also, the total population in the system is given by $\sum_{i=1}^n N_i^r$ and is therefore also positive and bounded.
\end{proof}

\subsection{Equilibrium of the mobility model}
\label{subsec:EPmigration}
To further analyze system \eqref{sys:dN}, the following notation is introduced.
Let $\ba=(a_1,\ldots,a_n)\in\IR^n$ be a vector. 
The notation $\ba\geq\bZero$ indicates that $a_i\geq 0$ for all $i=1,\ldots,n$; $\ba>\bZero$ means that $\ba\geq\bZero$ and there exists $i\in\{1,\ldots,n\}$ such that $a_i>0$; finally, $\ba\gg\bZero$ indicates that $\ba$ is entry-wise positive.
The same notation is used for matrices.
Here and later, it is convenient to use the following vector notation for state variables.
For $X=N$ and, later, $X\in\{N,S,I,R,V\}$, denote
\begin{subequations}\label{eq:ordering_X}
\begin{align}
\bX_i &= (X_{i1},\ldots,X_{in})^T \\
\begin{split}
\bm{X} &= (\bm{X}_1^T,\ldots,\bm{X}_n^T)^T \\
&= (X_{11},\ldots,X_{1n},X_{21},\ldots,X_{nn})^T.
\end{split}
\end{align}
\end{subequations}
Thus, $\bN_i$ is the part of $\bN$ corresponding to the subpopulation residing in patch $i$.
Let $\M=\oplus_{i=1}^n\M_{i}$ be the block diagonal
mobility matrix, with each block $\mathcal{M}_{i}$
describing the movement of residents of patch $i$ and given by
\begin{equation}\label{eq:M}
\M_{i}=\begin{pmatrix}
-\sum_{\substack{k=1\\ k\neq 1}}^nm_{ik1} & m_{i12} & \cdots & m_{i1n} \\
m_{i21} & -\sum_{\substack{k=1\\ k\neq 2}}^nm_{ik2} & \cdots & m_{i2n} \\
&&& \\
m_{in1} & m_{in2} & & -\sum_{\substack{k=1\\ k\neq n}}^nm_{ikn}
\end{pmatrix}.
\end{equation}
By \cite[Lemma 2]{ArinoBajeuxKirkland2019}, $-\M_{i}$ is a singular M-matrix.
From now on, assume that each matrix $\M_{i}$ is irreducible, i.e., the mobility digraph for residents of patch $i$ is strongly connected (if residents of patch $i$ only visit a subset of patches, $\M_i$ is restricted to that accessible subset, with $N_{ij}\equiv 0$ for unvisited patches $j$).
It is not assumed that all $\M_i$ have the same zero/nonzero pattern, just that they all be irreducible on their respective accessible sets.
Let
\[
\D_i=\diag(d_{i1},\ldots,d_{in})
\]
and
\[
\D = \bigoplus_{i=1}^n\D_i
\]
be the diagonal matrices of death rates. Finally, let $\bb=(\bb_{1}^T,\ldots,\bb_{n}^T)^T$, with the vector $\bb_i$ having $i$th entry $\birth_{ii}$ and zeros everywhere else.
The vector $\bb$ is the nonhomogeneous component of the system and system \eqref{sys:dN} can be written as 
\begin{equation}\label{sys:dN_B}
\frac{d}{dt}\bN=(\mathcal{M}-\D)\bN+\bb.
\end{equation}
Because of the structure of the mobility model used, individuals resident in a given patch ``radiate'' out to all the patches to which they can travel, but never lose their connection to their patch of residence. 
As a consequence, the resident population of patch $i$ satisfies the equation
\begin{equation}\label{eq:dNi}
\frac{d}{dt}\bN_i=(\mathcal{M}_{i}-\D_i)\bN_i+\bb_i
\end{equation}
and thus can be considered independently for each $i=1,\ldots,n$.
Since $\D_i>\bZero$, this model has a unique equilibrium as given
by the following result.
\begin{proposition}\label{prop:EPmigration}
The mobility model \eqref{sys:dN} always has a unique asymptotically stable equilibrium $\bN^\star\gg\bZero$, in which the equilibrium value of the resident population of patch $i=1,\ldots,n$ is given by $\bN_i^\star=(\D_i-\M_{i})^{-1}\bb_i$.
\end{proposition}
\begin{proof}
From \eqref{eq:dNi}, the value of the equilibrium is given by $\bN_i^\star=(\D_i-\mathcal{M}_{i})^{-1}\bb_i$ provided that the matrix $\D_i-\M_{i}$ is invertible.
Since $\D_i>\bZero$ and $\M_{i}$ is irreducible, it follows from \cite[Proposition 3(4)]{ArinoBajeuxKirkland2019} that $\D_i-\M_i$ is an irreducible nonsingular M-matrix and $(\D_i-\M_i)^{-1}\gg\bZero$. 
As a consequence, $\bN_i^\star=(\D_i-\M_{i})^{-1}\bb_i$ is unique and $\bN_i^\star\gg \bZero$ since $\bb_i>\bZero$.
Finally, since $\D_i-\M_{i}$ is an irreducible nonsingular M-matrix, it has all its eigenvalues with positive real parts, implying that $\M_i-\D_i$ has all its eigenvalues with negative real parts and $\bN_i^\star$ is asymptotically stable. 
\end{proof}


\section{The SIRS model}
\label{sec:SIRS_model}
Before considering the model with vaccination, properties of the underlying SIRS epidemic model are briefly summarised.
To the best of the authors' knowledge, properties of an SIRS model in a resident-traveller context have not been established. 

\subsection{Formulation of the model}
\label{sec:SIRS-model-formulation}
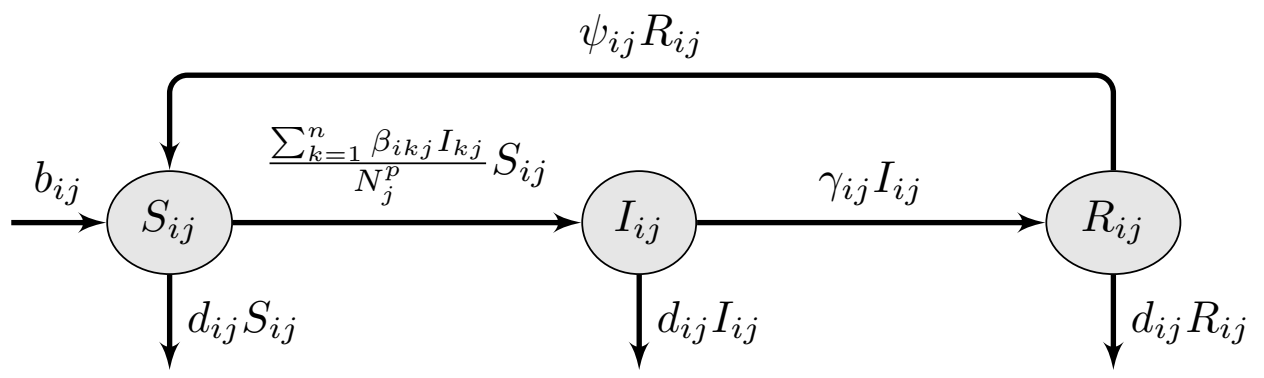
\begin{figure}[htbp]
    \centering
    \resizebox{\linewidth}{!}{%
    \begin{tikzpicture}[auto, 
	cloud/.style={minimum width={width("N-1")+2pt},
		draw, ellipse,fill=gray!20}]
	\node [cloud] (S) {$S_{ij}$};
	\node [cloud, right=2.75cm of S] (I) {$I_{ij}$};
	\node [cloud, right=2.75cm of I] (R) {$R_{ij}$};
    \node [cloud, left=0.75cm of S, draw=none, fill=none] (b) {};
    \node [cloud, below=0.75cm of S, draw=none, fill=none] (dS) {};
    \node [cloud, below=0.75cm of I, draw=none, fill=none] (dI) {};
    \node [cloud, below=0.75cm of R, draw=none, fill=none] (dR) {};
	\path [line, very thick] (S) to node [midway, above] (TextNode) {$\frac{\sum_{k=1}^n\beta_{ikj}I_{kj}}{N_j^p}S_{ij}$} (I);
	\path [line, very thick] (I) to node [midway, above] (TextNode) {$\gamma_{ij} I_{ij}$} (R);
	\path [line, very thick, rounded corners] (R.north) -- ++(0,0.75) coordinate (topR1) -- (topR1 -| S.north) node [midway, above] (TextNode) {$\psi_{ij} R_{ij}$} -- (S.north);
	\path [line, very thick] (b) to node [midway, above] (TextNode) {$b_{ij}$} (S);
	\path [line, very thick] (S) to node [midway, right] (TextNode) {$d_{ij}S_{ij}$} (dS);
	\path [line, very thick] (I) to node [midway, right] (TextNode) {$d_{ij}I_{ij}$} (dI);
	\path [line, very thick] (R) to node [midway, right] (TextNode) {$d_{ij}R_{ij}$} (dR);
    \end{tikzpicture}%
    }
\caption{Flow diagram for one subpopulation of the SIRS model in a given population $N_{ij}$. There is birth into the $S_{ij}$ compartment with $i=j$, i.e., $b_{ij}=0$ if $i\neq j$.}
\label{fig:SIRS}
\end{figure}

The following assumptions are made.
\begin{enumerate}
\item
Each subpopulation is described by a compartmental SIRS model; the model structure in each subpopulation is as in Figure~\ref{fig:SIRS}.
Here, $S$, $I$ and $R$ are the numbers of individuals that are susceptible, infectious and recovered, respectively.
The number of infectious residents in patch $i$ present in patch $j$ at time $t$ is denoted by $I_{ij}(t)$, and similarly for the other variables. The average duration of infectiousness for an individual resident in patch $i$ currently in patch $j$ is $1/\gamma_{ij}$, and upon recovery the average period of temporary immunity is $1/\psi_{ij}$.
\item
The disease transmission coefficient in patch $j$ from an infectious resident in patch $k$ to a susceptible resident in patch $i$ is denoted $\beta_{ikj}\geq 0$. It is assumed that $\sum_{j=1}^n\beta_{ikj}>0$, i.e., the disease can be transmitted in some patch $j$ from residents of patch $k$ to individuals from patch $i$ visiting patch $j$. Incidence is supposed to be of standard type.
\item 
As in the underlying mobility model of Section~\ref{sec:mobility}, birth only occurs in the patch of residence and happens at the rate $\birth_{ij}$, with $\birth_{ij}=0$ for $i\neq j$. 
There is no vertical transmission of the disease, so all individuals are born susceptible.
\item The rate of movement can depend on the epidemiological status of individuals.
Using the same notation as in Section~\ref{sec:mobility}, a superscript is added to indicate the status of individuals, so movement is described by
\[
m_{\textrm{residence},\textrm{destination},\textrm{origin}}^\textrm{status}.
\]
This defines a multi-digraph with at most $3n$ arcs in each direction between any two vertices. 
Unless otherwise indicated, it is assumed that the matrices $\M_{i}^X-\D_i$ are irreducible, where $\M_{i}^X$ is given for $X\in\{S,I,R\}$ by \eqref{eq:M} with $m_{ijk}$ replaced by $m_{ijk}^X$.
\item There is no disease induced mortality: diseases such as poliomyelitis, although debilitating, rarely cause death.
\end{enumerate}

For $i,j=1,\ldots,n$, the equations for this $n$-patch model are
\begin{subequations}\label{sys:nnSIRS}
\begin{align}
\begin{split}
\frac{d}{dt}S_{ij} &= 
\birth_{ij}+\psi_{ij} R_{ij} 
-\left(d_{ij}+\frac{\sum_{k=1}^n\beta_{ikj}I_{kj}}{N_j^p}\right)S_{ij} \\
&\quad +\sum_{k=1}^n m_{ijk}^S S_{ik} 
\end{split}
\label{sys:nnSIRS_dSij} \\
\begin{split}
\frac{d}{dt}I_{ij} &= 
\frac{\sum_{k=1}^n\beta_{ikj}I_{kj}}{N_j^p}S_{ij}
-\left(d_{ij}+\gamma_{ij}\right)I_{ij} \\
&\quad +\sum_{k=1}^n m_{ijk}^II_{ik}
\end{split}
\label{sys:nnSIRS_dIij} \\
\frac{d}{dt}R_{ij} &= \gamma_{ij}I_{ij}
-\left(d_{ij}+\psi_{ij}\right)R_{ij}
+\sum_{k=1}^n m_{ijk}^RR_{ik}
\label{sys:nnSIRS_dRij}
\end{align}
\end{subequations}
System \eqref{sys:nnSIRS} is considered with initial conditions
\[
\sum_{j=1}^nS_{ij}(0)>0, \quad
I_{ij}(0)\geq 0
\quad\text{and}\quad
R_{ij}(0)\geq 0
\]
for all $i,j=1,\ldots,n$ and can easily be shown to have all solutions in the nonnegative orthant.
It is also easily shown that the set
\[
\{
(S_{ij},I_{ij},R_{ij})\in\IR^{3n^2}; \forall i,j=1,\ldots,n,\ I_{ij}=0
\}
\]
is positively invariant under the flow of \eqref{sys:nnSIRS}.
From now on, it is further assumed that
\[
\sum_{i,j=1}^n I_{ij}(0)>0
\]
to avoid trivial solutions.

It is convenient to use vector and matrix notation as in Section~\ref{subsec:EPmigration}, so for $X\in\{N,S,I,R\}$, the state variable vectors are written using \eqref{eq:ordering_X}.
Using the same ordering of first and second index, write $\Psi=\diag(\psi_{ij})$ the matrix of rates of loss of immunity and $\Gamma=\diag(\gamma_{ij})$ the matrix of rates of recovery.
When used with only one index, variables refer to matrices relative to residents of a given patch. For instance, $\Gamma_i=\diag(\gamma_{i1},\ldots,\gamma_{in})$, so $\Gamma=\bigoplus_{i=1}^n\Gamma_i$.

\subsection{The disease-free equilibrium of the SIRS model}
\label{subsec:DFE_SIRS}
In order to continue the analysis, system \eqref{sys:nnSIRS} is investigated at the disease-free equilibrium (DFE), \emph{i.e.}, where $I_{ij}=0$ for all $i,j=1,\ldots,n$.
In earlier studies \cite{ArinoVdD2003b,ArinoVdD2003a}, the case of status-independent mobility was considered, in which case the population $N_{ij}$ in subpopulation $(i,j)$ tends to $N_{ij}^\star$ as given in Proposition~\ref{prop:EPmigration}.
Because of status dependent travel, the situation is more complex here. First, at the DFE, there are no recovered individuals.
\begin{lemma}\label{lemma:Rgoesto0_SIRS}
Suppose that system \eqref{sys:nnSIRS} is at the DFE. 
Then $\lim_{t\to\infty}\bR(t)=\bm{0}$.
\end{lemma}
\begin{proof}
Substituting $I_{ij}=0$ in \eqref{sys:nnSIRS_dRij} gives
\begin{equation}
\frac{d}{dt}R_{ij} = \sum_{k=1}^n m_{ijk}^RR_{ik}
-(d_{ij}+\psi_{ij})R_{ij},
\end{equation}
or, in vector form,
\[
\frac{d}{dt}\bR=(\mathcal{M}^R-\D-\Psi)\bR.
\]
The matrix $-\mathcal{M}^R$ is a singular M-matrix, so reasoning as in the proof of Proposition~\ref{prop:EPmigration}, $\D+\Psi-\M^R$ is a nonsingular M-matrix. 
As a consequence, $\bR=\bm{0}$ is the unique equilibrium of $\bR'=(\M^R-\D-\Psi)\bR$, $\M^R-\D-\Psi$ has all eigenvalues with negative real parts and therefore, $\bR\to\bm{0}$ as $t\to\infty$. 
\end{proof}

It follows that at the DFE,
\begin{align*}
0 &= \birth_{ij}-d_{ij}S_{ij} +\sum_{k=1}^n m_{ijk}^SS_{ik}.
\end{align*}
In vector notation, these equations take the form, for patch $i=1,\ldots,n$,
\begin{equation}\label{sys:nnSIRS_DFE}
\bm{0} = (\mathcal{M}^S_i-\D_i)\bS_i+\bb_i.
\end{equation}
From this, the situation can be determined at the level of the patch of residence.
\begin{lemma}\label{lemma:DFE_SIRS_patch_constant_birth}
Consider residents of patch $i=1,\ldots,n$. 
If system \eqref{sys:nnSIRS} is at the DFE, then the DFE for residents of patch $i$ is unique and given by
\begin{equation}\label{eq:DFE_SIRS}
\bS_i^\star=(\D_i-\M^S_{i})^{-1}\bb_i\gg\bm{0},\quad \bI_i^\star=\bR_i^\star=\bm{0}.
\end{equation}
\end{lemma}
\begin{proof}
Consider the DFE relation \eqref{sys:nnSIRS_DFE}.
The matrix $\D_i-\M_i^S$ is a nonsingular M-matrix and by the irreducibility assumption, has entry-wise positive inverse \cite[Proposition 3(4)]{ArinoBajeuxKirkland2019}.
Solving \eqref{sys:nnSIRS_DFE} gives $\bS_i^\star$ as in \eqref{eq:DFE_SIRS}; its entry-wise positivity follows from the inverse positivity of $\D_i-\M_i^S$.
\end{proof}

Note that if $\M^S=\M$, this is the equilibrium given by Proposition~\ref{prop:EPmigration} for the demographic model; at the DFE \eqref{eq:DFE_SIRS}, the total subpopulation sizes are given by $\bN_i^\star=\bS_i^\star$.

\subsection{The basic reproduction number $\R_0$}
\label{subsec:R0}
The next generation operator approach \cite{DiekmannHeesterbeek2000} and its characterisation in the ordinary differential equations case given by \cite{VdDWatmough2002} is used to determine the basic reproduction number $\R_0$ for the system.
The reasoning closely follows that used in \cite{ArinoVdD2003a}. 
Consider only the infected variables and order them as in \eqref{eq:ordering_X}. 
Let the vectors $\mathcal{F}$ and $-\mathcal{W}$ contain new infections and all other flows within infected compartments, respectively.
Differentiating the vectors $\mathcal{F}$ and $\mathcal{W}$ with respect to the infected variables and evaluating at the disease-free equilibrium \eqref{eq:DFE_SIRS} gives the matrices
\begin{equation}
\label{eq:F_matrix_SIRS}
F=\left(
\begin{matrix}
F_{11} & \cdots & F_{1n} \\
\vdots && \vdots \\
F_{n1} & \cdots & F_{nn}
\end{matrix}
\right)
\end{equation}
and
\begin{equation}
\label{eq:W_matrix_SIRS}
W=\Gamma+\D-\M^I
=\bigoplus_{i=1}^n\left(
\Gamma_i+\D_i-\M_i^I
\right).
\end{equation}
The blocks $F_{ij}$ in matrix $F$ are the Jacobian matrices obtained by differentiating $\mathcal{F}_i$ with respect to $\bI_j$.
They are diagonal and take the form
$F_{ij}=\diag(f_{ij1},\ldots,f_{ijn})$, with
\begin{equation}
\label{eq:Fij_blocks_SIRS}
f_{ijq}=
\beta_{ijq}
\frac{S^\star_{iq}}{N_q^{p\star}}=
\beta_{ijq}
\frac{N_{iq}^\star}{N_q^{p\star}},
\end{equation}
where $q=1,\ldots,n$ and $N_q^{p\star}=\sum_{k=1}^nN_{kq}^\star$ is the total population of patch $q$ at the DFE.
Hence $F$ is a nonnegative matrix. 
Also, $W$ is a nonsingular M-matrix and thus $W^{-1}$ is nonnegative. 
In fact,
\begin{equation}
\label{eq:inv_W_matrix_SIRS}
W^{-1}
=\bigoplus_{i=1}^n(\Gamma_i+\D_i-\M_i^I)^{-1}
=:\bigoplus_{i=1}^nW^{-1}_{ii}
\end{equation}
and the $W_{ii}^{-1}$ are positive by the irreducibility assumption.
As a consequence, $FW^{-1}$ takes the form
\begin{equation}
\label{eq:FWinv_SIRS}
FW^{-1}=\left(
\begin{matrix}
F_{11}W^{-1}_{11} & \cdots & F_{1n}W^{-1}_{nn} \\
\vdots & & \vdots \\
F_{n1}W^{-1}_{11} & \cdots & F_{nn}W^{-1}_{nn}
\end{matrix}
\right),
\end{equation}
and \cite[Theorem 2]{VdDWatmough2002} gives the following result, in which $\rho(\cdot)$ is the spectral radius, showing that the basic reproduction number $\R_0$ acts as a threshold.
\begin{theorem}\label{th:R0_nSIRS}
Consider system \eqref{sys:nnSIRS} with $F$ and $W$ as defined above. Let 
\begin{equation}\label{eq:R0}
\R_0=\rho(FW^{-1}).
\end{equation}
Then
\begin{itemize}
\item if $\R_0<1$, the DFE is locally asymptotically stable;
\item if $\R_0>1$, the DFE is unstable, the disease is uniformly persistent and there exists at least one endemic equilibrium point.
\end{itemize}
\end{theorem}

\begin{proof}
The local asymptotic stability for $\R_0<1$ and instability for $\R_0>1$ follow from \cite[Theorem 2]{VdDWatmough2002}. 
It remains to show uniform persistence and existence of an endemic equilibrium when $\R_0>1$. 
Let $\Omega$ be the biologically feasible region defined by the non-negative orthant of the state space where $N_i^r > 0$ for all $i=1,\ldots,n$. 
Because the total population in each patch is governed by a dissipative demographic equation, solutions are bounded, making $\Omega$ positively invariant and point dissipative.
Let $\Omega_0$ be the subset of $\Omega$ where $I_{ij} > 0$ for at least one pair $(i,j)$ and define the disease-free subspace as $\partial\Omega_0 = \Omega\setminus\Omega_0$, where $\bI =\bm{0}$. 

The DFE is the unique equilibrium in $\partial\Omega_0$ and is globally asymptotically stable within $\partial\Omega_0$ by Lemmas~\ref{lemma:Rgoesto0_SIRS} and \ref{lemma:DFE_SIRS_patch_constant_birth}. 
When $\R_0 > 1$, the spectral bound $s(F-W) > 0$. 
It is now shown that the DFE is a weak repeller for $\Omega_0$. 
Assume by contradiction that there exists a solution starting in $\Omega_0$ such that $\limsup_{t\to\infty} \bI(t) = \bm{0}$. 
Then $\bR(t)\to\bm{0}$ and $\bS(t)\to\bS^\star$ as $t\to\infty$, so that $S_{iq}(t)/N_q^p(t)\to S_{iq}^\star/N_q^{p\star}$. 
Consequently, for any $\eta > 0$ there exists $t_1 > 0$ such that for all $t > t_1$, $\bI'(t) \ge (F - W - \eta M) \bI(t)$ for some non-negative matrix $M$. 
By the continuity of the spectral bound, $\eta$ can be chosen small enough such that $s(F - W - \eta M) > 0$. 
This implies that $\bI(t)$ must eventually grow exponentially, which contradicts the assumption that $\bI(t) \to \bm{0}$. 
Therefore, the DFE repels solutions starting in $\Omega_0$.

Because the system is point dissipative, $\Omega$ is positively invariant and $\partial\Omega_0$ contains only the isolated and acyclic invariant set $\{E_0\}$ (the DFE), it follows from standard uniform persistence theory \cite[Theorem 4.6]{Thieme1993} that the disease is uniformly persistent, which together with the convexity of $\Omega$ implies the existence of at least one endemic equilibrium point in $\Omega_0$.
\end{proof}

\subsection{Basic reproduction number in isolated patches}
\label{sec:R0_SIRS_isolation}
The basic reproduction number in isolated patches is useful in numerical considerations and is easily obtained from the previous reasoning. 
Suppose that the patches are isolated, i.e., $\M^S=\M^I=\M^R=\bZero$. 
Then the model can be considered with no subpopulations, as in the absence of movement, the number of visitors to a patch is either identically zero or goes to zero.
Using \eqref{eq:DFE_SIRS} with $\M_{i}^S=0$, the DFE takes the form, for $i=1,\ldots,n$,
\[
(S_{ii}^\star,I_{ii}^\star,R_{ii}^\star) = 
\left(
\frac{b_{ii}}{d_{ii}},0,0
\right).
\]
Matrix $F$ reduces to the scalar
\[
F_{ii}=\beta_{iii}
\frac{S^\star_{ii}}{N^\star_{ii}}
=\beta_{iii},
\]
while $W_{ii}^{-1}$ is
\[
W_{ii}^{-1} = \frac{1}{\gamma_{ii}+d_{ii}},
\]
so the basic reproduction number in patch $i$ in isolation is
\[
\R_{0i}^{\eqref{sys:nnSIRS}}=
\frac{\beta_{iii}}{\gamma_{ii}+d_{ii}}.
\]

\subsection{Case of movement equal for all epidemiological states}
\label{sec:equal_mvt_SIRS}

\begin{lemma}
\label{lemma:behaviour_total_pop_epidemic_model}
Suppose that in system \eqref{sys:nnSIRS}, individuals in all disease states travel at the same rate, i.e., $\M:=\M^S=\M^I=\M^R$. 
Then the number of individuals $N_{ij}$ in each subpopulation $(i,j)$ is convergent.
\end{lemma}
\begin{proof}
Denote $m_{ijk}:=m_{ijk}^S= m_{ijk}^I= m_{ijk}^R$. 
For a given subpopulation $(i,j)$,
\[
\frac{d}{dt}N_{ij}
=\frac{d}{dt}(S_{ij}+I_{ij}+R_{ij})
=\birth_{ij}-d_{ij}N_{ij}+\sum_{k=1}^n m_{ijk}N_{ik}.
\]
This is equation \eqref{sys:dN} and thus Proposition~\ref{prop:EPmigration} holds here and $N_{ij}$ converges for all $i,j=1,\ldots,n$. 
As previously, this limit is denoted $N^\star_{ij}$.
\end{proof}

This result is similar to those obtained in the SIS model in \cite{ArinoVdD2003a} and the SEIRS model in \cite{ArinoVdD2003b}, where movement is equal for all compartments.
Because the $N_{ij}$ converge, it is possible to use a comparison theorem for asymptotically autonomous limit systems to prove that the DFE is globally asymptotically stable; see Appendix~\ref{app:DFE_SIRS_GAS}.

\section{The model with vaccination}
\label{sec:epidemic_model}
The model in the presence of a disease for which a vaccine is available is now considered. 
\subsection{Formulation of the model with vaccination}
\label{sec:SVIRS-formulation}
\begin{figure}[htbp]
    \centering
    \resizebox{\linewidth}{!}{%
    \begin{tikzpicture}[auto, 
	cloud/.style={minimum width={width("N-1")+2pt},
		draw, ellipse,fill=gray!20}]
	\node [cloud] (S) {$S_{ij}$};
	\node [cloud, below right=2cm of S] (V) {$V_{ij}$};
	\node [cloud, right=2.75cm of S] (I) {$I_{ij}$};
	\node [cloud, right=2.75cm of I] (R) {$R_{ij}$};
    \node [cloud, left=2cm of S, draw=none, fill=none] (bS) {};
    \node [cloud, left=2cm of V, draw=none, fill=none] (bV) {};
    \node [cloud, below=0.75cm of S, draw=none, fill=none] (dS) {};
    \node [cloud, above=0.75cm of I, draw=none, fill=none] (dI) {};
    \node [cloud, below=0.75cm of R, draw=none, fill=none] (dR) {};
    \node [cloud, below=0.75cm of V, draw=none, fill=none] (dV) {};
	\path [line, very thick] (S) to node [midway, below] (TextNode) {$\frac{\sum_{k=1}^n\beta_{ikj}I_{kj}}{N_j^p}S_{ij}$} (I);
	\path [line, very thick] (V) to node [midway, right] (TextNode) {$\frac{\sum_{k=1}^n\beta_{ikj}I_{kj}}{N_j^p}\pi_{ij}V_{ij}$} (I);
	\path [line, very thick] (S) to node [midway, below, sloped] (TextNode) {$\phi_{ij}S_{ij}$} (V);
	\path [line, very thick] (I) to node [midway, below] (TextNode) {$\gamma_{ij} I_{ij}$} (R);
	\path [line, very thick, rounded corners] (R.north) -- ++(0,1.25) coordinate (topR2) -- (topR2 -| S.north) node [midway, above] (TextNode) {$\psi_{ij} R_{ij}$} -- (S.north);
	\path [line, very thick] (bS) to node [midway, above] (TextNode) {$(1-p_{ij})b_{ij}$} (S);
	\path [line, very thick] (bV) to node [midway, below] (TextNode) {$p_{ij}b_{ij}$} (V);
	\path [line, very thick] (S) to node [midway, left] (TextNode) {$d_{ij}S_{ij}$} (dS);
	\path [line, very thick] (I) to node [midway, right] (TextNode) {$d_{ij}I_{ij}$} (dI);
	\path [line, very thick] (R) to node [midway, right] (TextNode) {$d_{ij}R_{ij}$} (dR);
	\path [line, very thick] (V) to node [midway, right] (TextNode) {$d_{ij}V_{ij}$} (dV);
    \end{tikzpicture}%
    }
\caption{Flow diagram for one subpopulation of the SVIRS model with vaccination in a given population $N_{ij}$. There is birth into the S and V compartments when $i=j$, i.e., $b_{ij}=0$ if $i\neq j$.}
\label{fig:SVIRS}
\end{figure}
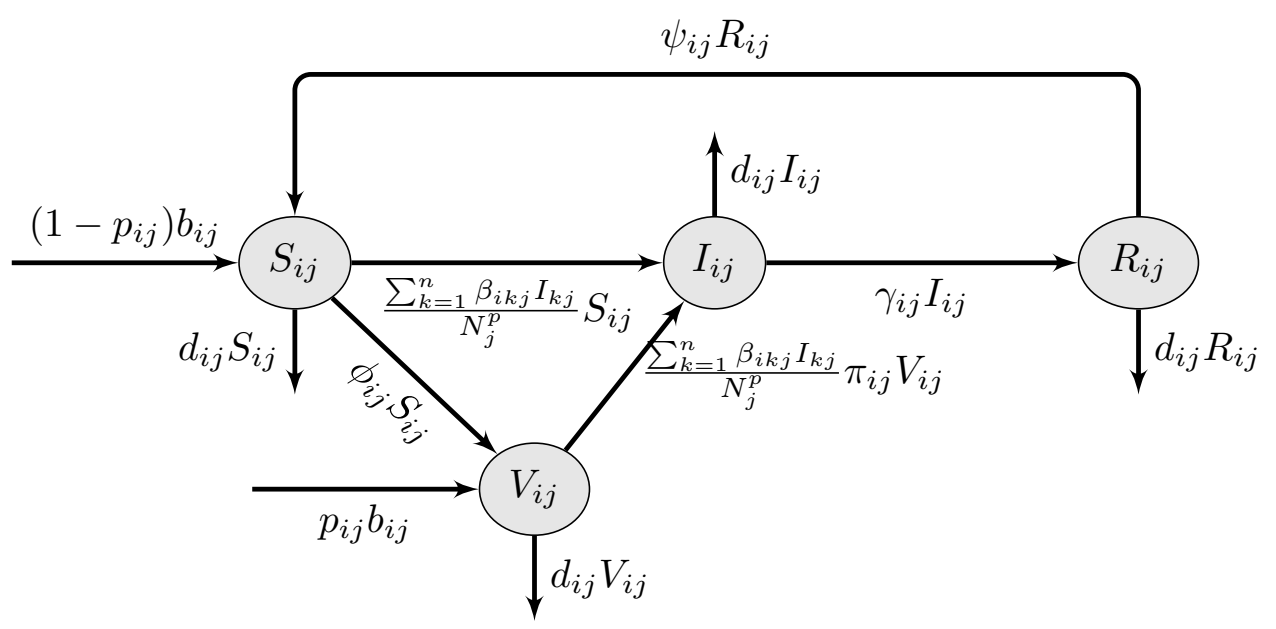

The following assumptions are made additionally to those in Section~\ref{sec:SIRS_model}.
\begin{enumerate}
\item
Each subpopulation is described by a compartmental SIRS model with vaccination, \emph{i.e.}, an SVIRS model; the model structure in each subpopulation is as in Figure~\ref{fig:SVIRS}. 
Here, $S$, $I$, $R$, $V$ are the numbers of individuals that are susceptible, infectious, recovered, vaccinated, respectively. 
The number of vaccinated residents in patch $i$ present in patch $j$ at time $t$ is denoted by $V_{ij}(t)$, and similarly for the other variables.
\item 
A proportion $p_{ij}$ (with $0\leq p_{ij}<1$) of newborns in subpopulation $(i,j)$ are vaccinated at birth (since birth only occurs in the patch of residence, this only applies when $i=j$), and susceptible individuals resident in patch $i$ are vaccinated at the rate $\phi_{ij}$ when they are in patch $j$. 
Vaccination reduces disease susceptibility by a factor $\pi_{ij}$ with $0\leq \pi_{ij}\leq 1$, thus $1-\pi_{ij}$ is the vaccine efficacy in subpopulation $(i,j)$.
Thus, in this population, $\pi_{ij}=0$ means the vaccine is completely effective at preventing infection, whereas $\pi_{ij}=1$ means that the vaccine is useless. Once infected, vaccinated and unvaccinated individuals transmit the disease at the same rate.
\end{enumerate}

Note that the vaccination model used here is simple. In particular, omitting vaccine waning avoids the standard vaccine-waning mechanism for backward bifurcation in an isolated patch \cite{ArinoMccluskeyVdD2003} (and rules out backward bifurcation altogether when immunity is permanent, $\psi_{ij}=0$). Although backward bifurcations are known to occur in metapopulation epidemic models (see, e.g., \cite{ArinoDucrotZongo2012}), the point here would be less clear if a backward bifurcation were to occur.

For $i,j=1,\ldots,n$, the equations for this $n$-patch model are
\begin{subequations}\label{sys:nnSVIRS}
\begin{align}
\begin{split}
\frac{d}{dt}S_{ij} &= 
(1-p_{ij})\birth_{ij}+\psi_{ij} R_{ij} +\sum_{k=1}^n m_{ijk}^S S_{ik} \\
&\quad -\left(d_{ij}+\phi_{ij}
+\frac{\sum_{k=1}^n\beta_{ikj}I_{kj}}{N_j^p}\right)S_{ij}
\end{split}
\label{sys:nnSVIRS_dSij} \\
\begin{split}
\frac{d}{dt}I_{ij} &= 
\left(S_{ij}+\pi_{ij} V_{ij}\right)
\frac{\sum_{k=1}^n\beta_{ikj}I_{kj}}{N_j^p}
-\left(d_{ij}+\gamma_{ij}\right)I_{ij} \\
&\quad +\sum_{k=1}^n m_{ijk}^II_{ik}
\end{split}
\label{sys:nnSVIRS_dIij} \\
\frac{d}{dt}R_{ij} &= \gamma_{ij}I_{ij}
-\left(d_{ij}+\psi_{ij}\right)R_{ij}
+\sum_{k=1}^n m_{ijk}^RR_{ik}
\label{sys:nnSVIRS_dRij} \\
\begin{split}
\frac{d}{dt}V_{ij} &= p_{ij}\birth_{ij} + \phi_{ij}S_{ij} 
-\left(d_{ij}
+\pi_{ij}\frac{\sum_{k=1}^n\beta_{ikj}I_{kj}}{N_j^p}
\right)V_{ij} \\
&\quad +\sum_{k=1}^n m_{ijk}^VV_{ik}.
\end{split}
\label{sys:nnSVIRS_dVij}
\end{align}
\end{subequations}
As with the SIRS model \eqref{sys:nnSIRS}, if there are initially no infectious individuals in the entire system, then this remains true for all times. 
As a consequence, system \eqref{sys:nnSVIRS} is considered with initial conditions 
\[
\sum_{j=1}^nS_{ij}(0)>0, \
\sum_{i,j=1}^n I_{ij}(0)>0, \
R_{ij}(0)\geq 0
\ \text{and}\ 
V_{ij}(0)\geq 0
\]
for all $i,j=1,\ldots,n$ and can easily be shown to have all solutions in the nonnegative orthant.

As earlier, it is convenient to use vector and matrix notation, so for $X\in\{S,I,R,V\}$, \eqref{eq:ordering_X} is used.
Using the same ordering of first and second index, write $\Phi=\diag(\phi_{ij})$ the matrix of vaccination rates, $\Gamma=\diag(\gamma_{ij})$ the matrix of rates of recovery and
$\Pi=\diag(\pi_{ij})$ the matrix of vaccine-induced reduction of transmissibility.
When used with only one index, variables refer to matrices relative to residents of a given patch. For instance, $\Phi_i=\diag(\phi_{i1},\ldots,\phi_{in})$, so $\Phi=\bigoplus_{i=1}^n\Phi_i$.

\subsection{The disease-free equilibrium of the SVIRS model}
\label{subsec:DFE_SVIRS}
In order to continue the analysis, system \eqref{sys:nnSVIRS} is investigated at the disease-free equilibrium (DFE), \emph{i.e.}, where $I_{ij}=0$ for all $i,j=1,\ldots,n$.
In earlier studies \cite{ArinoVdD2003b,ArinoVdD2003a}, the case of status-independent mobility was considered, in which case the population $N_{ij}$ in subpopulation $(i,j)$ tends to $N_{ij}^\star$ as given in Proposition~\ref{prop:EPmigration}.
Because of vaccination and status dependent travel, the situation is more complex here. 
First, similarly to the SIRS model of Section~\ref{sec:SIRS_model}, at the DFE there are no recovered individuals, since Lemma~\ref{lemma:Rgoesto0_SIRS} holds true for \eqref{sys:nnSVIRS} without any change to the proof.
It follows that at the DFE,
\begin{align*}
0 &= (1-p_{ij})\birth_{ij}-\left(d_{ij}+\phi_{ij}\right)S_{ij} +\sum_{k=1}^n m_{ijk}^SS_{ik}\\
0 &= p_{ij}\birth_{ij} + \phi_{ij}S_{ij}-d_{ij}V_{ij}+\sum_{k=1}^n m_{ijk}^VV_{ik}.
\end{align*}
In vector notation, these equations take the form, for patch $i=1,\ldots,n$,
\begin{equation}\label{eq:nnSVIRS_DFE_Bconstant}
\begin{pmatrix}0\\0\end{pmatrix} 
=
\begin{pmatrix}
\mathcal{M}^S_{i}-\Phi_i-\D_i & 0 \\
\Phi_i & \mathcal{M}^V_{i}-\D_i 
\end{pmatrix}
\begin{pmatrix} \bS_i\\\bV_i \end{pmatrix}
+\begin{pmatrix}(\mathbb{I}-P_i)\bb_i\\ P_i\bb_i\end{pmatrix},
\end{equation}
where $P_i = \diag(p_{i1},\ldots,p_{in})$ is the diagonal matrix of vaccination at birth proportions for residents of patch $i$.
From this, the situation at the patch level can be determined.
\begin{lemma}\label{lemma:DFE_patch_constant_birth}
For system \eqref{sys:nnSVIRS}, the DFE of residents in patch $i=1,\ldots,n$ is unique and given by
\begin{equation}\label{eq:DFE_SVIRS}
\begin{gathered}
\bS_i^\star=(\Phi_i+\D_i-\M^S_{i})^{-1}(\mathbb{I}-P_i)\bb_i\gg 0,\quad \bI_i^\star=\bR_i^\star=0,\\
\bV_i^\star=(\D_i-\M^V_{i})^{-1}(\Phi_i\bS_i^\star + P_i\bb_i).
\end{gathered}
\end{equation}
\end{lemma}
\begin{proof}

Since $\Phi_i+\D_i>0$, $\D_i>0$ and $\M_i^S,\M_i^V$ irreducible, it follows from \cite[Proposition 3(4)]{ArinoBajeuxKirkland2019} that $\Phi_i+\D_i-\M_{i}^S$ and $\D_i-\M_{i}^V$ are nonsingular M-matrices with entry-wise positive inverses.
Solving \eqref{eq:nnSVIRS_DFE_Bconstant} at the equilibrium gives $\bS_i^\star$ and $\bV_i^\star$ as in \eqref{eq:DFE_SVIRS}. Here, $\bS_i^\star$ is entry-wise positive (since $p_{ii}<1$) and $\bV_i^\star$ is entry-wise positive provided that $\Phi_i$ has a positive entry or $p_{ii}>0$.
\end{proof}

\subsection{The vaccination reproduction number $\R_v$}
\label{subsec:Rv}
The reasoning here closely follows that used in Section~\ref{subsec:R0}.
Matrix $F$ has the same block structure as \eqref{eq:F_matrix_SIRS} with blocks $F_{ij}$ taking the form $F_{ij}=\diag(f_{ij1},\ldots,f_{ijn})$, with
\begin{equation}
\label{eq:Fij_blocks_SVIRS}
f_{ijq}=\beta_{ijq}
\frac{S^\star_{iq}+\pi_{iq}V^\star_{iq}}{N^{p\star}_{q}},
\end{equation}
where $N_q^{p\star} = \sum_{\ell=1}^n (S_{\ell q}^\star + V_{\ell q}^\star)$, $q=1,\ldots,n$.
The following is the pendant of Theorem~\ref{th:R0_nSIRS}.
\begin{theorem}\label{th:R0_nnSVIRS}
Consider system \eqref{sys:nnSVIRS} with $F$ and $W$ defined by \eqref{eq:F_matrix_SIRS} and \eqref{eq:W_matrix_SIRS}, the blocks in \eqref{eq:F_matrix_SIRS} having entries of the form \eqref{eq:Fij_blocks_SVIRS}. 
Let the system-wide vaccination reproduction number be
\begin{equation}\label{eq:Rv_nnSVIRS}
\R_v=\rho(FW^{-1}).
\end{equation}
Then
\begin{itemize}
\item if $\R_v<1$, the DFE is locally asymptotically stable;
\item if $\R_v>1$, the DFE is unstable, the disease is uniformly persistent, and there exists at least one endemic equilibrium point.
\end{itemize}
\end{theorem}

\begin{proof}
The local stability of the DFE follows directly from \cite[Theorem 2]{VdDWatmough2002}. Furthermore, because the state space remains positively invariant, point dissipative, and the demographic processes bound the population, the proof of uniform persistence and the existence of an endemic equilibrium point follows the same arguments as those detailed in the proof of Theorem~\ref{th:R0_nSIRS} via standard uniform persistence theory \cite[Theorem 4.6]{Thieme1993}.
\end{proof}

As in Section~\ref{sec:R0_SIRS_isolation}, it is useful to compute the vaccination number that holds when patches are isolated from each other, i.e., when there is no movement.
This value is indeed used in numerical simulations to set some of the parameters.
Suppose that patches are isolated; as in Section~\ref{sec:R0_SIRS_isolation}, the model can be considered with no subpopulations, as in the absence of movement, the number of visitors to a patch is either identically zero or goes to zero.
Using \eqref{eq:DFE_SVIRS} with $\M_{i}^S=\M_{i}^V=0$, the DFE takes the form
\[
(S_{ii}^\star,I_{ii}^\star,R_{ii}^\star,V_{ii}^\star) = 
\left(
\frac{(1-p_{ii})b_{ii}}{d_{ii}+\phi_{ii}},0,0,\frac{b_{ii}(p_{ii}d_{ii}+\phi_{ii})}{d_{ii}(d_{ii}+\phi_{ii})}
\right).
\]
The matrix $F$ reduces to the scalar
\[
F_{ii}=\beta_{iii}
\frac{S^\star_{ii}+\pi_{ii}V^\star_{ii}}{N^\star_{ii}},
\]
while $W_{ii}^{-1}$ is
\[
W_{ii}^{-1} = \frac{1}{\gamma_{ii}+d_{ii}},
\]
so the vaccination reproduction number in patch $i$ in isolation is
\begin{equation}
\label{eq:Rv_isolation_SVIRS}
\R_{vi}^{\eqref{sys:nnSVIRS}}=
\frac{\beta_{iii}}{\gamma_{ii}+d_{ii}}
\;
\frac{S^\star_{ii}+\pi_{ii}V^\star_{ii}}{N_{ii}^\star}.
\end{equation}

As in the case of the SIRS model, when movement is the same for all epidemiological states, the following result holds.
\begin{lemma}
\label{lemma:behaviour_total_pop_SVIRS}
If individuals in all disease states travel at the same rate, i.e., $\M:=\M^S=\M^I=\M^R=\M^V$, then the number of individuals $N_{ij}$ in each subpopulation $(i,j)$ is convergent.
\end{lemma}
\begin{proof}
Denote $m_{ijk}:=m_{ijk}^S= m_{ijk}^I= m_{ijk}^R= m_{ijk}^V$. For a given subpopulation $(i,j)$,
\[
\frac{d}{dt}N_{ij}=\frac{d}{dt}(S_{ij}+I_{ij}+R_{ij}+V_{ij})
=\birth_{ij}-d_{ij}N_{ij}+\sum_{k=1}^n m_{ijk}N_{ik}.
\]
This is equation \eqref{sys:dN} and thus Proposition~\ref{prop:EPmigration} holds here and $N_{ij}$ converges for all $i,j=1,\ldots,n$. As previously, this limit is denoted $N^\star_{ij}$.
\end{proof}

\subsection{The high-mobility limit}
\label{subsec:high-mobility-limit}

Part of the objective of this paper is to consider the differences between the resident-traveller and classic metapopulation models. 
In order to do so, the classic SVIRS model in Appendix~\ref{app:classic-SVIRS} is used.
Basic properties of that model are given in the appendix, with, in particular, the expressions of the reproduction numbers $\R_0^\eqref{sys:nSVIRS}$ and $\R_v^\eqref{sys:nSVIRS}$. 
The relation between both model types is investigated computationally in Section~\ref{sec:special_cases}.

However, before proceeding further, the distinction between the two modelling paradigms is formalised by considering the analytical limit of the vaccination reproduction number as the mobility rate $m \to \infty$.
Let the movement matrix be parameterized as $\M_c = m \bar{\M}_c$, where $\bar{\M}_c$ defines the digraph topology.

In the classic metapopulation framework, the spectral radius of the next-generation matrix $\R_v^\eqref{sys:nSVIRS} = \rho(F_c W_c^{-1})$ depends on the transition matrix $W_c = \Gamma_c + \D_c - m\bar{\M}_c$. 
Assuming that disease-independent mortality and recovery rates are uniform across patches ($d_i=d$, $\gamma_i=\gamma$), it follows that $\Gamma_c + \D_c = (d+\gamma)\mathbb{I}$. 
By factoring out $m$, the inverse is written as $W_c^{-1} = \frac{1}{m} \left( \frac{d+\gamma}{m} \mathbb{I} - \bar{\M}_c \right)^{-1}$. 
Because $\bar{\M}_c$ defines a strongly connected movement digraph (its columns sum to zero), it has a simple eigenvalue at $0$ with right eigenvector $\bu$ (the stationary distribution, normalised so that $\bm{1}^T\bu = 1$) and left eigenvector $\bm{1}^T$.
Using singular perturbation theory for Markov chains (see, e.g., \cite[Theorem 2.1]{YinZhang1998} or \cite[Chapter II, Equation (1.14)]{Kato1995}), as $m \to \infty$ the inverse $W_c^{-1}$ converges to a rank-1 multiple of a projection onto the null space of $\bar{\M}_c$, given by $\frac{1}{d+\gamma} \bu \bm{1}^T$.
Furthermore, as $m \to \infty$, the disease-free equilibrium states perfectly mix across the network. The proportion of susceptible and vaccinated individuals in every patch converges to the global network averages, denoted $\bar{s}^\star$ and $\bar{v}^\star$. Therefore, the transmission matrix $F_c(m)$ converges to a fully homogenized matrix $\bar{F}_c = \diag\left( \beta_i (\bar{s}^\star + \pi_i \bar{v}^\star) \right)$.

Consequently, the next generation matrix $F_c W_c^{-1}$ converges to a rank-1 matrix, and its spectral radius converges to the trace of this limiting matrix:
\begin{equation}
\begin{split}
    \lim_{m \to \infty} \R_v^\eqref{sys:nSVIRS} &= \rho\left( \frac{1}{d+\gamma} \bar{F}_c \bu \bm{1}^T \right) = \text{tr}\left( \frac{1}{d+\gamma} \bar{F}_c \bu \bm{1}^T \right) \\
    &= \sum_{i=1}^n u_i \frac{\beta_i}{d+\gamma} (\bar{s}^\star + \pi_i \bar{v}^\star).
\end{split}
\end{equation}
Thus, infinite mobility in the classic model completely homogenizes the system. 
The system-wide reproduction number becomes a weighted arithmetic mean of the transmission potentials, driven entirely by the globally averaged vaccine coverage $\bar{v}^\star$. 
If a single patch completely halts vaccination, its true isolated potential exceeds unity ($\R_{vi}^{\eqref{sys:nSVIRS}} > 1$). However, if the surrounding patches are highly vaccinated, the globally averaged susceptibility ($\bar{s}^\star + \pi_i \bar{v}^\star$) can still fall to a level where the trace drops below unity, erroneously allowing analytical eradication in the classic model.

Conversely, in the resident-traveller framework, the population does not assimilate into destination cohorts. 
Residents of an unvaccinated patch $i$ ($\Phi_i = 0$, $P_i = 0$) distribute across the digraph according to their specific stationary visiting distribution $\bu^{(i)}$, but remain completely unvaccinated ($\bS_i^\star = \bN_i^\star$, $\bV_i^\star = \bZero$) regardless of how heavily other patches vaccinate their own residents. 
Because the next-generation matrix $FW^{-1}$ tracks infections by residency, the principal submatrix $F_{ii}W_{ii}^{-1}$ corresponding to transmission \emph{exclusively among} unvaccinated residents of patch $i$ (or, more generally, the principal submatrix corresponding to the set of all non-vaccinating patches) remains strictly positive. 
By the Perron-Frobenius theorem (see, e.g., \cite[Theorem 8.1.18]{HornJohnson1990}), the spectral radius of the full system is bounded below by the spectral radius of any principal submatrix:
\begin{equation}
    \mathcal{R}_v \ge \rho(F_{ii}W_{ii}^{-1}),
\end{equation}
where $F_{ii}W_{ii}^{-1}$ represents transmission among the susceptible residents of patch $i$. 
Whenever this submatrix spectral radius (or the spectral radius of the subnetwork of unvaccinated patches) exceeds unity, global eradication is impossible regardless of the vaccination rate in the vaccinating patches, in contrast to the classic model where high vaccination in some patches can drive the system-wide reproduction number below unity.
Note that, unlike the classic result, this lower bound is not a limit as $m\to\infty$: it holds for every movement rate.
As $m$ increases, residents of patch $i$ spend a growing fraction of their time in other patches, where they make up a smaller share of the population present, so the entries of $F_{ii}$ decrease and $\rho(F_{ii}W_{ii}^{-1})$ can fall below unity (see Section~\ref{subsec:vacc-allbut1patch}).
The bound is therefore most informative at low to moderate mobility.

\section{Investigation of specific vaccination situations}
\label{sec:special_cases}
A few special cases are now considered assuming different types of vaccination patterns in the patches. 
These scenarios are investigated analytically or numerically, with a focus on comparing the outcomes predicted by the resident-traveller model versus the classic metapopulation model. 

\subsection{Setting of the numerical work}
In the cases where numerical results are sought, three graph architectures are considered.
\begin{enumerate}
    \item \textbf{Complete graph --} In this situation, each vertex is connected to the $n-1$ other vertices.
    \item \textbf{Lattice --} In this case, patches are laid out on a lattice and residents of a given patch are only able to travel to patches immediately ``north'', ``south'', ``east'' and ``west'' of their patch of residence.
    A torus is considered, with the ``top'' connecting to the ``bottom'' and ``right'' connecting to the ``left''.
    This is an approximation of a two-dimensional diffusion process with periodic boundary conditions.
    \item \textbf{Ring graph --} Patches form a ring, with access to exactly the two adjacent patches.
    This is an approximation of a one-dimensional diffusion process with periodic boundary conditions.
\end{enumerate}
Note that only graph architectures that are regular, i.e., have an equal number of neighbours for all patches, are considered. This avoids artificial demographic shifts in the classic metapopulation model where patches with more neighbours would accumulate individuals when a uniform total leaving rate is divided equally among outgoing edges. The simple lattice with reflecting boundaries is therefore omitted from theoretical comparisons, though non-equal degrees will be considered in the real-world example.

\begin{table}[htbp]
    \centering
    \resizebox{\linewidth}{!}{%
    \begin{tabular}{c|c|c}
        Parameter & Interpretation & Baseline value \\
        \hline
        $1/\gamma_{ij}$ & Duration infection & 21 days \\
        $1/\psi_{ij}$ & Duration immune period & 1 year \\
        $\phi_{ij}$ & Vaccination rate & $10^{-5}$--$10^{-2}\text{ day}^{-1}$ (or \emph{computed}) \\
        $1-\pi_{ij}$ & Vaccine efficacy & 0.7 (70\%) \\
        $c^\star_{ij}$ & Vaccine coverage at DFE & 0.8 (80\%, reference value) \\
        $b_{ii}$ & Birth rate for residents of $i$ & \emph{Computed} ($d_{ii} N_i^r$) \\
        $1/d_{ij}$ & Life expectancy & 70 years \\
        $N_i^r$ & Resident population of patch $i$ & 10,000 \\
        $m$ & Total outbound movement rate & $10^{-3}$--$10^{-1}\text{ day}^{-1}$ \\
        $r$ & Return rate to home patch & $0.09\text{ day}^{-1}$ \\
        $\R_{0i}^{\eqref{sys:nnSIRS}}$ & Isolated basic reproduction number & 1.2 \\
        $\R_{vi}^{\eqref{sys:nnSVIRS}}$ & Isolated vaccination reproduction number (at $c^\star=0.8$) & 0.528 (reference value)
    \end{tabular}%
    }
    \caption{Parameters and values in the model and baseline values used in simulations. \emph{Computed} values are deduced from values of other parameters. The coverage $c^\star$ and the corresponding isolated $\R_{vi}^{\eqref{sys:nnSVIRS}}$ are reference values illustrating the effect of vaccination in an isolated patch; in the simulations on synthetic networks, the vaccination rate $\phi$ is varied directly over the range indicated.}
    \label{tab:parameters}
\end{table}

Baseline values of parameters are generally chosen as in Table~\ref{tab:parameters}.
In all simulations on the synthetic networks ($n=49$ patches), infectious individuals are assumed to travel at $10\%$ of the rate of susceptible, recovered and vaccinated individuals ($\M^I = 0.1\M^S$, $\M^R=\M^V=\M^S$; similarly $\M_c^I = 0.1\M_c^S$ in the classic model).
Values of the vaccination rate $\phi_{ij}$ and birth rate $b_{ii}$ are computed from assumptions about quantities they determine, as explained now.

\paragraph{Vaccine coverage}
Obtaining realistic parameter values for the vaccination rate is difficult.
Indeed, the model does not include age classes, which are the points in life history with known vaccination action. Instead, a continuous vaccination rate $\phi_{ij}$ is incorporated alongside a birth vaccination proportion $p_{ij}$.
Vaccination applies to the entire susceptible population, such as may approximate the vaccination for measles following an outbreak.
The vaccination rate is then set indirectly, using an observable consequence, namely, the vaccine coverage, i.e., the proportions $c_{ij}=V_{ij}/N_{ij}\in[0,1]$ of individuals in the population that are vaccinated.
This is particularly relevant since data are usually given in terms of vaccine coverage.
At the DFE \eqref{eq:DFE_SVIRS}, this can be written as
\begin{equation}\label{eq:vaccine_coverage}
c_{ij}^\star=
\frac{V_{ij}^\star}{S_{ij}^\star+V_{ij}^\star}
\end{equation}
but this estimation only accurately reflects the true vaccination coverage in a completely susceptible population, which does not hold when an endemic state emerges for $\R_v>1$.
Also, even when $\R_v<1$, \eqref{eq:DFE_SVIRS} involves $\phi_{ij}$ through a nontrivial matrix inverse, so obtaining $\Phi$ using \eqref{eq:vaccine_coverage} involves a difficult inverse problem.
Since, in the numerics, vaccination occurs only in the home patch and no vaccination at birth is assumed ($P_i = 0$), the expression for \eqref{eq:vaccine_coverage} can be simplified using \eqref{eq:DFE_SVIRS}, giving that at the DFE, $\bV_i^\star = (\D_i-\M_i^V)^{-1}\Phi_i\bS_i^\star$. Using the diagonal structure of $\Phi_i$ and dividing numerator and denominator by $S_{ii}^\star$ gives
\[
c_{ii}^\star
=\frac{\left[(\D_i-\M_i^V)^{-1}\right]_{ii}\phi_{ii}}
{1+\left[(\D_i-\M_i^V)^{-1}\right]_{ii}\phi_{ii}}.
\]
This implies that complete vaccine coverage $c_{ij}^\star=1$ can never be achieved, except in the limit as $\phi_{ii}\to\infty$. 
More generally, whenever vaccination at birth is not complete ($p_{ij} < 1$), all non-vaccinated newborns spend some time, however short, in the susceptible compartment.

\paragraph{Setting birth rates --}
The birth rate $\bb$ is chosen such that the demographic component plays a minimal role during simulations and each patch maintains a target equilibrium resident population $N_i^{r\star}$. 
In the resident-traveller model, births occur only in the home patch ($b_{ij}=0$ for $i\neq j$). With uniform per capita death rate $d_{ij}=d$, setting $b_{ii}=d N_i^{r\star}$ determines the unique demographic equilibrium $\bN_i^\star=(\D_i-\M_i)^{-1}\bb_i$, which satisfies $\sum_{j=1}^n N_{ij}^\star = N_i^{r\star}$. 
In the absence of disease ($\bI=\bR=\bZero$) and with $\M^S=\M^V=\M$, the DFE components satisfy $\bS_i^\star+\bV_i^\star=\bN_i^\star$, so that starting with $\bN(0)=\bN^\star$, the subpopulation sizes remain constant at the demographic level. 
When movement of infectious individuals differs from that of other states ($\M^I\neq\M^S$) so that Lemma~\ref{lemma:behaviour_total_pop_epidemic_model} does not hold during an outbreak, there are only small transient variations of the spatial distribution $\bN_i(t)$ while the total resident population $N_i^r(t)=N_i^{r\star}$ remains strictly constant.

\paragraph{Vaccination only in the patch of residence}
\label{subsec:vacc-residence-patch}
Although the model allows for vaccination in all subpopulations, it is supposed in numerical work that individuals are vaccinated only when present in their place of residence, \emph{i.e.}, $\phi_{ij}=0$ for $i\neq j$.
Then the matrix $\Phi$ takes the form
\[
\tilde\Phi=\diag(\tilde\Phi_1,\ldots,\tilde\Phi_n),
\]
where $\tilde\Phi_i$ is the matrix with the only nonzero entry $(i,i)$ equal to $\phi_{ii}$.
Consider the $i$th diagonal block $(\D_i-\M_i^V)^{-1}$ in $(\D-\M^V)^{-1}$ and let $[(\D_i-\M_i^V)^{-1}]_{ij}$ be the $(i,j)$ entry in that matrix.
Because of the structure of $\Phi_i$, the matrix $(\D_i-\M_i^V)^{-1}\Phi_i$ has only one nonzero column that consists of $\phi_{ii}$ times the entries in the $i$th column of $(\D_i-\M_i^V)^{-1}$ and thus $\bV_i^\star$ takes the form
\begin{equation}
\label{eq:Vstar_vacc_home_patch}
\bV_i^\star = 
\left(
\left[(\D_i-\M_i^V)^{-1}\right]_{1i},
\ldots,
\left[(\D_i-\M_i^V)^{-1}\right]_{ni}
\right)^T\ \phi_{ii}S_{ii}^\star.
\end{equation}

Note that this assumption is realistic from a public health perspective.
Routine childhood immunisation schedules and publicly funded vaccination programmes are typically administered through local health systems and registries tied to an individual's place of residence, rather than to temporary visitors.

\subsection{Vaccination in only one patch}
\label{subsec:vacc_1patch}
Suppose that vaccination takes place only in one patch. 
To consider implications on the model, assume without loss of generality that the patch vaccinating is patch 1.
In this case, $\Phi=\diag(\Phi_1,0,\ldots,0)$ and $P_i=0$ for $i=2,\ldots,n$.
Using this in \eqref{eq:DFE_SVIRS}, the DFE takes the form
\begin{align*}
\bS_1^\star &= (\Phi_1+\D_1-\M^S_{1})^{-1}(\mathbb{I}-P_1)\bb_1, \quad \bI_1^\star=\bR_1^\star=\bZero, \\
\bV_1^\star &= (\D_1-\M^V_{1})^{-1}(\Phi_1\bS_1^\star+P_1\bb_1),
\end{align*}
and, for $i=2,\ldots,n$,
\[
\bS_i^\star = (\D_i-\M_i^S)^{-1}\bb_i,\quad \bI_i^\star=\bR_i^\star=\bV_i^\star=\bZero.
\]
Thus, $\bV^\star=\left({\bV_1^\star}^T,0,\ldots,0\right)^T$ and diagonal entries in $F_{ij}$ take the form \eqref{eq:Fij_blocks_SVIRS} when $i=1$ and \eqref{eq:Fij_blocks_SIRS} when $i=2,\ldots,n$.

Numerically, the case investigated has patch 1 vaccinating (which can be thought of as the patch in position (1,1) in a torus grid) and is such that, in isolation, $\R_{v1}^{\eqref{sys:nnSVIRS}}<1$, while none of the others vaccinate and all have $\R_{vi}^{\eqref{sys:nnSVIRS}}>1$ for $i \neq 1$. This setting addresses the first question posed in the Introduction: can patch 1 contribute to disease eradication in the other patches, and can it protect its own population?

In the classic metapopulation framework \eqref{sys:nSVIRS}, the vaccination matrix takes the form $\Phi_c = \diag(\phi_1, 0, \ldots, 0)$. Using \eqref{eq:DFE_classic}, the number of vaccinated individuals at the disease-free equilibrium is given by $\bV_c^\star=(\D_c-\M_c)^{-1}\Phi_c\bS_c^\star$. Since $\D_c - \M_c$ is an irreducible, non-singular M-matrix for a strongly connected digraph, its inverse is strictly positive. Consequently, the product $(\D_c-\M_c)^{-1}\Phi_c\bS_c^\star$ yields a vector where every single entry is strictly positive. This means that at the DFE, every patch's resident population in the classic model becomes partially vaccinated because visitors to patch 1 are vaccinated and subsequently assimilated into other patches. This alters the basic reproduction number computation for all patches globally. In contrast, the resident-traveller model ensures $\bV_i^\star = 0$ for all $i \neq 1$, preserving the fully susceptible status of residents of non-vaccinating patches.

To visualize this divergence, a parameter sweep of the system-wide reproduction number $\R_v$ is performed across varying movement rates ($m$) and vaccination rates ($\phi$) for the scenario where only one patch vaccinates. For the remaining 48 untreated patches in the $7\times 7$ grid, the isolated local reproduction number is configured to $\R_0 = 1.2$. 

In the resident-traveller formulation, vaccination is strictly governed by residence. Thus, individuals from the 48 untreated patches are not subject to vaccination when travelling to the vaccinating patch. Because these untreated patches remain fully susceptible reservoirs, the overall $\R_v$ is analytically bounded below by the spectral radius of the unvaccinated subnetwork, remaining approximately constant at $\R_v \approx 1.20$. Consequently, the answer to the first part of the question is negative in the resident-traveller framework: for the parameter range considered, it is impossible to eradicate the disease globally by vaccinating only a single patch.

This numerical observation can be formalized analytically. In the resident-traveller formulation, a lack of vaccination in a specific jurisdiction inherently places a lower bound on the system-wide reproduction number, which prevents eradication whenever this bound exceeds unity. Let $\widetilde{\mathcal{R}}_0^k = \rho(F_{kk} W_{kk}^{-1})$ be defined as the digraph-adjusted basic reproduction number for residents of patch $k$ in the absence of vaccination. This reflects transmission within the interconnected system, as opposed to the fully isolated $\mathcal{R}_0^k$ defined in Section~\ref{sec:R0_SIRS_isolation}.

\begin{theorem}
Assume vaccination only occurs in the patch of residence ($\phi_{ij}=0$ for $i \neq j$). 
Let patch $k$ be a non-vaccinating patch, meaning there is no ongoing vaccination ($\phi_{kk} = 0$) and no vaccination at birth ($p_{kk} = 0$). 
Then the system-wide vaccination reproduction number satisfies $\mathcal{R}_v \ge \widetilde{\mathcal{R}}_0^k$. 
\end{theorem}

\begin{proof}
Because patch $k$ does not vaccinate its residents, the $k$-th block of the ongoing vaccination matrix is $\Phi_k = 0$. Furthermore, since there is no vaccination at birth for residents of patch $k$, the birth vaccination proportion matrix satisfies $P_k\bb_k = 0$. 
It follows that any initially vaccinated residents of this patch eventually die, i.e., $\bV_k \to \bZero$ at the DFE.

From Lemma~\ref{lemma:DFE_patch_constant_birth}, the equilibrium vaccinated population for residents of patch $k$ at the disease-free equilibrium is given by $\bV_k^\star = (\D_k-\M^V_{k})^{-1}(\Phi_k\bS_k^\star + P_k\bb_k)$. Substituting $\Phi_k = 0$ and $P_k\bb_k = 0$ yields $\bV_k^\star = 0$, meaning residents of patch $k$ remain fully susceptible ($\bS_k^\star = \bN_k^\star$). 

Consequently, the $k$-th diagonal block $F_{kk}$ of the transmission matrix $F$ has entries 
\[
f_{kkq} = \beta_{kkq} \frac{S_{kq}^\star + \pi_{kq} V_{kq}^\star}{N_q^{p\star}} = \beta_{kkq} \frac{N_{kq}^\star}{N_q^{p\star}}.
\]
This matches the localized transmission block in the complete absence of vaccination. 

The global next-generation matrix is $F W^{-1}$. Because $W^{-1}$ is block-diagonal with respect to the patch of residence, the non-negative matrix $F W^{-1}$ contains the principal submatrix $F_{kk} W_{kk}^{-1}$, which isolates transmission exclusively among residents of patch $k$. 

By the Perron-Frobenius theorem and standard properties of non-negative matrices, the spectral radius of a matrix is bounded below by the spectral radius of any of its principal submatrices. Therefore, 
\[
\mathcal{R}_v = \rho(F W^{-1}) \ge \rho(F_{kk} W_{kk}^{-1}) = \widetilde{\mathcal{R}}_0^k.\qedhere
\]
\end{proof}

\begin{figure}[htbp]
    \centering
    \includegraphics[width=\linewidth]{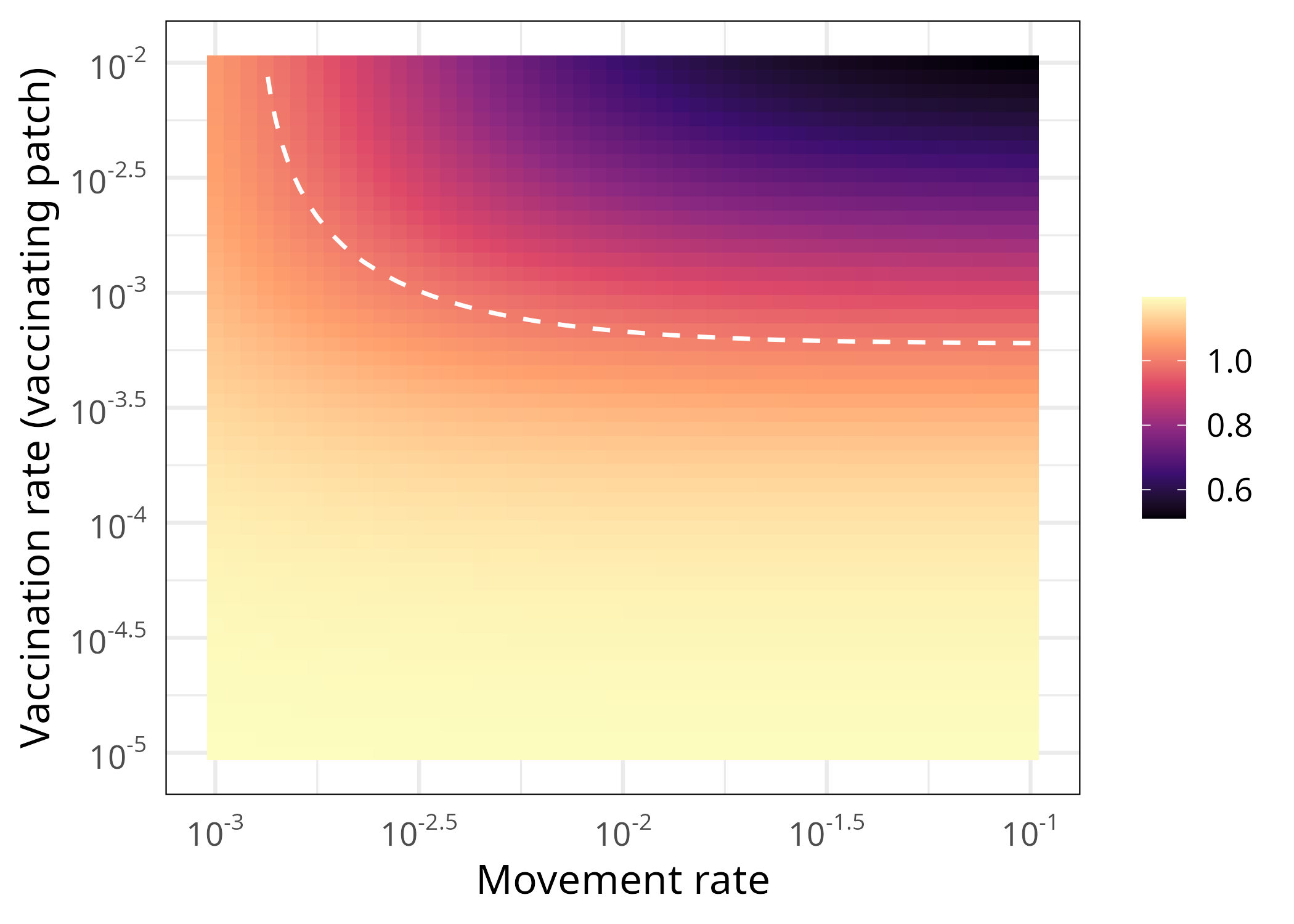}
    \caption{System-wide reproduction number $\R_v$ for the classic metapopulation model when only one patch vaccinates. The dashed contour line indicates the critical eradication threshold ($\R_v = 1$). The resident model (not shown) maintains a nearly constant $\R_v \approx 1.20$ across the same parameter space because untreated patches remain fully susceptible reservoirs.}
    \label{fig:heatmap_classic}
\end{figure}

Thus, if the local digraph-adjusted reproduction number of the untreated patch exceeds unity ($\widetilde{\mathcal{R}}_0^k > 1$), global eradication ($\mathcal{R}_v < 1$) is impossible, regardless of how heavily other patches vaccinate.
Regarding the second part of the question, vaccination in patch 1 only partially protects its own residents: while vaccinated residents of patch 1 have reduced susceptibility ($\pi_{11} < 1$), which lowers their infection rate compared to unvaccinated patches, imperfect vaccine efficacy ($\pi_{11} > 0$) and ongoing mixing with infectious visitors from the 48 non-vaccinating patches prevent patch 1 from completely shielding its population as long as the disease remains endemic in the network.
In contrast, as illustrated in Figure~\ref{fig:heatmap_classic}, the classic model applies vaccination by current physical location. When individuals from the untreated patches travel to the single vaccinating patch, they are vaccinated at its local rate and subsequently return or move on as vaccinated individuals. As the movement rate $m$ increases, this export of vaccinated individuals reduces the proportion of susceptibles across all patches, driving the global $\R_v$ below the eradication threshold ($\R_v < 1$).

\begin{figure*}[htbp]
    \centering
    \begin{minipage}{0.48\textwidth}
        \centering
        \includegraphics[width=\linewidth]{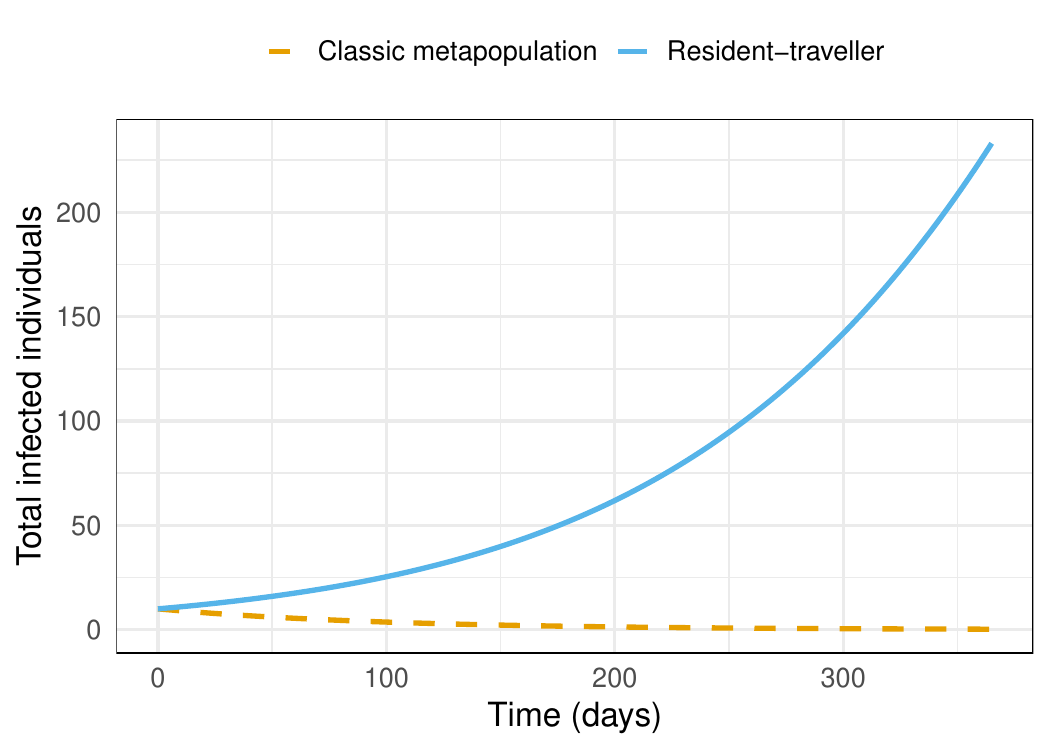}\\
        (a) Epidemic trajectories
    \end{minipage}\hfill
    \begin{minipage}{0.48\textwidth}
        \centering
        \includegraphics[width=\linewidth]{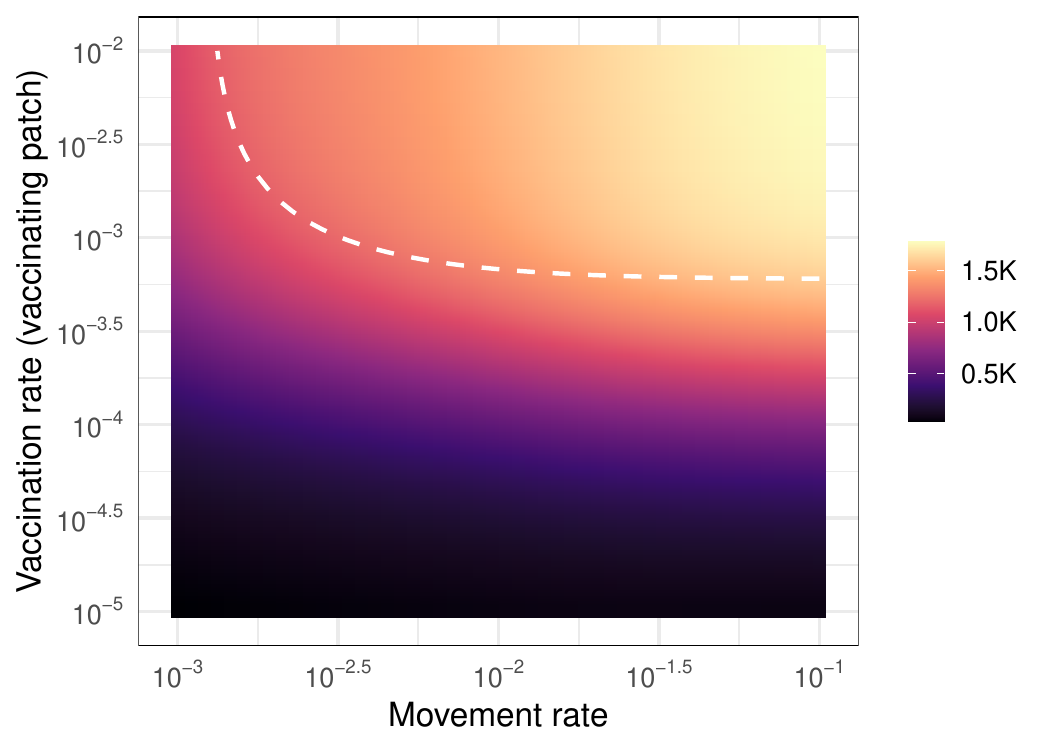}\\
        (b) Cumulative incidence difference
    \end{minipage}
    \caption{Epidemic curve time-series $I(t)$ for identical parameters ($m=0.01\text{ day}^{-1}$, $\phi_{11}=1/365\text{ day}^{-1}$) where $\R_v^\eqref{sys:nSVIRS} < 1 < \mathcal{R}_v^\eqref{sys:nnSVIRS}$ (a), and the corresponding difference in cumulative new infections between the resident-traveller and classic models over a one-year period (b).
The outbreak originates in an untreated patch in the $7\times 7$ grid.}
    \label{fig:vacc_one_patch_curves}
\end{figure*}

Figure~\ref{fig:vacc_one_patch_curves} shows the effect of this difference by plotting the epidemic curve $I(t)$ for identical parameters ($m = 0.01\text{ day}^{-1}$, $\phi_{11} = 1/365\text{ day}^{-1}$) where $\R_v^\eqref{sys:nSVIRS} < 1 < \mathcal{R}_v$, alongside the difference in cumulative incidence across the parameter space.
In this regime, the classic model predicts only a small transient wave before the outbreak dies out.
Conversely, the resident-traveller model resolves the large epidemic wave across the unprotected patches, resulting in substantially higher cumulative infections.

\subsection{Vaccination in all but one patch}
\label{subsec:vacc-allbut1patch}
Suppose that individuals are vaccinated in all patches except one.
Without loss of generality, assume that the non-vaccinating patch is patch $n$.
Patch $n$ could be the core of a city or region where individuals refuse to be vaccinated.
Patch $n$ does not vaccinate and has $\R_{vn}^{\eqref{sys:nnSVIRS}}>1$; all others vaccinate and have $\R_{vi}^{\eqref{sys:nnSVIRS}}<1$ for $i \neq n$. 
This scenario addresses the second question posed in the Introduction: what happens to the global system when a single jurisdiction ceases to vaccinate while surrounding locations continue to vaccinate?

For the classic metapopulation model, assuming patch $n$ does not vaccinate, the vaccination matrix takes the form $\Phi_c = \diag(\phi_1, \ldots, \phi_{n-1}, 0)$. 
Because individuals from patch $n$ travel to the other $n-1$ vaccinating patches, they become subject to vaccination there. 
Mathematically, the $n$-th component of the DFE vaccinated population $\bV_c^\star$ is given by the inner product of the $n$-th row of $(\D_c-\M_c)^{-1}$ and the vector $\Phi_c\bS_c^\star$. 
Due to the strict positivity of $(\D_c-\M_c)^{-1}$, this component is strictly positive. 
Consequently, the population of patch $n$ acquires vaccinated individuals through spatial mixing, which suppresses its effective susceptibility in the interconnected system. 
In the resident-traveller formulation, by contrast, residents of patch $n$ are not vaccinated when visiting other patches and remain a fully susceptible cohort.

\begin{figure*}[htbp]
    \centering
    \begin{subfigure}[b]{0.49\textwidth}
        \centering
        \includegraphics[width=\linewidth]{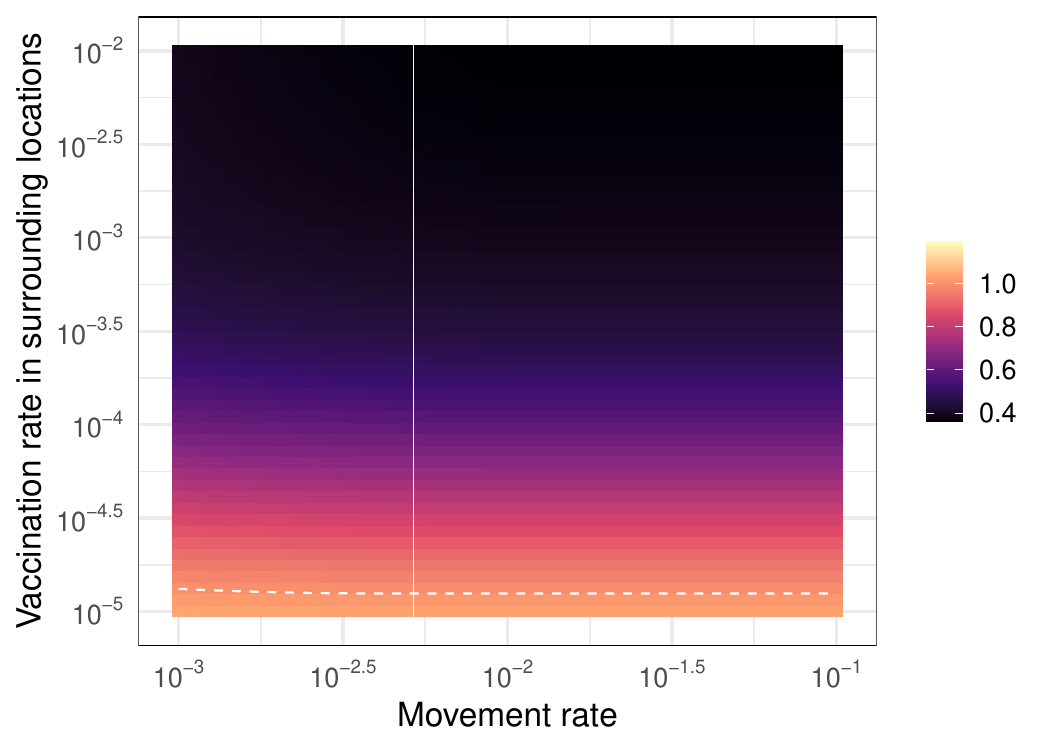}
        \caption{Classic metapopulation}
    \end{subfigure}
    \hfill
    \begin{subfigure}[b]{0.49\textwidth}
        \centering
        \includegraphics[width=\linewidth]{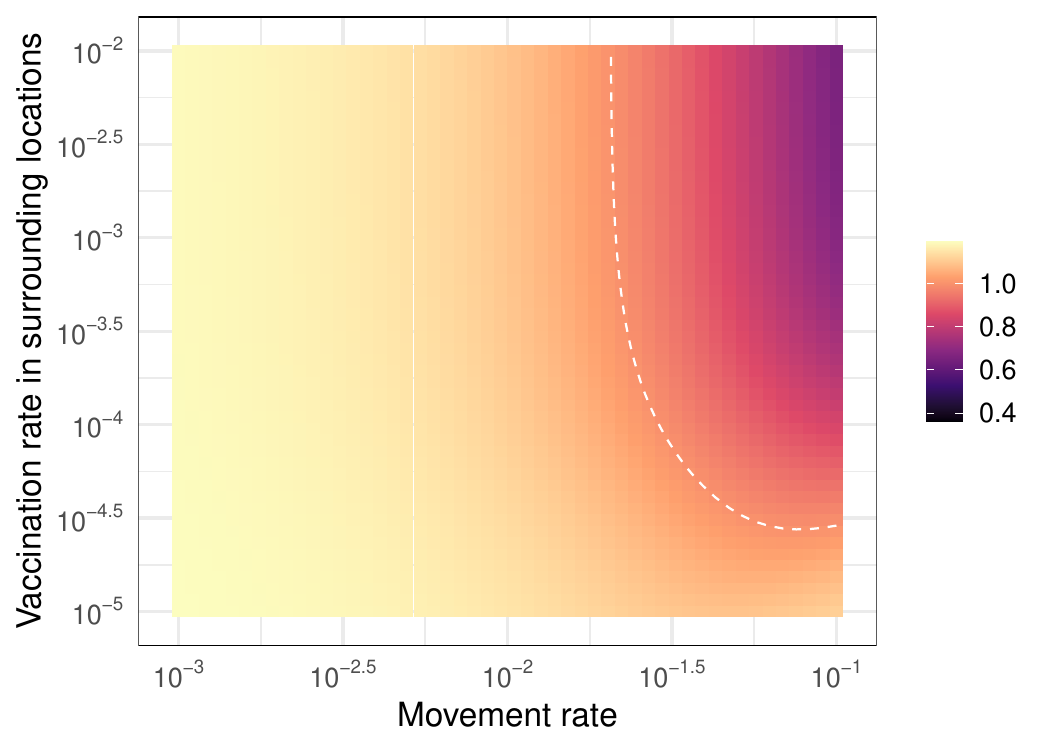}
        \caption{Resident-traveller}
    \end{subfigure}
    \caption{System-wide vaccination reproduction number $\mathcal{R}_v$ when a single core patch refuses to vaccinate ($\phi_n = 0$) while all surrounding locations vaccinate at rate $\phi$ in the classic model (a) and the resident-traveller model (b).
    The dashed white contour indicates the eradication threshold ($\mathcal{R}_v = 1$).}
    \label{fig:heatmap-ring-vaccination-rv}
\end{figure*}

Figure~\ref{fig:heatmap-ring-vaccination-rv} illustrates the system-wide reproduction number $\mathcal{R}_v$ across a parameter space of movement rates and vaccination rates in the surrounding locations for both models.
In the classic metapopulation model (left panel), even modest vaccination rates in the surrounding locations ($\phi \gtrsim 1.5\times 10^{-5}\text{ day}^{-1}$) create a ``safe'' region ($\mathcal{R}_v < 1$) across all movement rates, suggesting that immunizing surrounding patches easily compensates for a core patch's refusal to vaccinate.
However, the resident-traveller formulation (right panel) answers the second question from the Introduction differently: for low to moderate movement rates ($m \lesssim 0.02\text{ day}^{-1}$), where residents spend most of their time in their home patch so that $\widetilde{\mathcal{R}}_0^n > 1$, $\mathcal{R}_v$ strictly exceeds unity regardless of how heavily the surrounding locations vaccinate.
Only when the movement rate is sufficiently high ($m > 0.02\text{ day}^{-1}$) that residents of the unvaccinated core patch spend a large fraction of their time diluted among the vaccinated surrounding locations ($\widetilde{\mathcal{R}}_0^n < 1$) can high vaccination in those locations bring the system-wide $\mathcal{R}_v$ below unity.

\subsection{Further considerations on digraph topology}
\label{subsec:topology-considerations}
\begin{figure*}[htbp]
    \centering
    \begin{subfigure}[b]{0.49\textwidth}
        \centering
        \includegraphics[width=\linewidth]{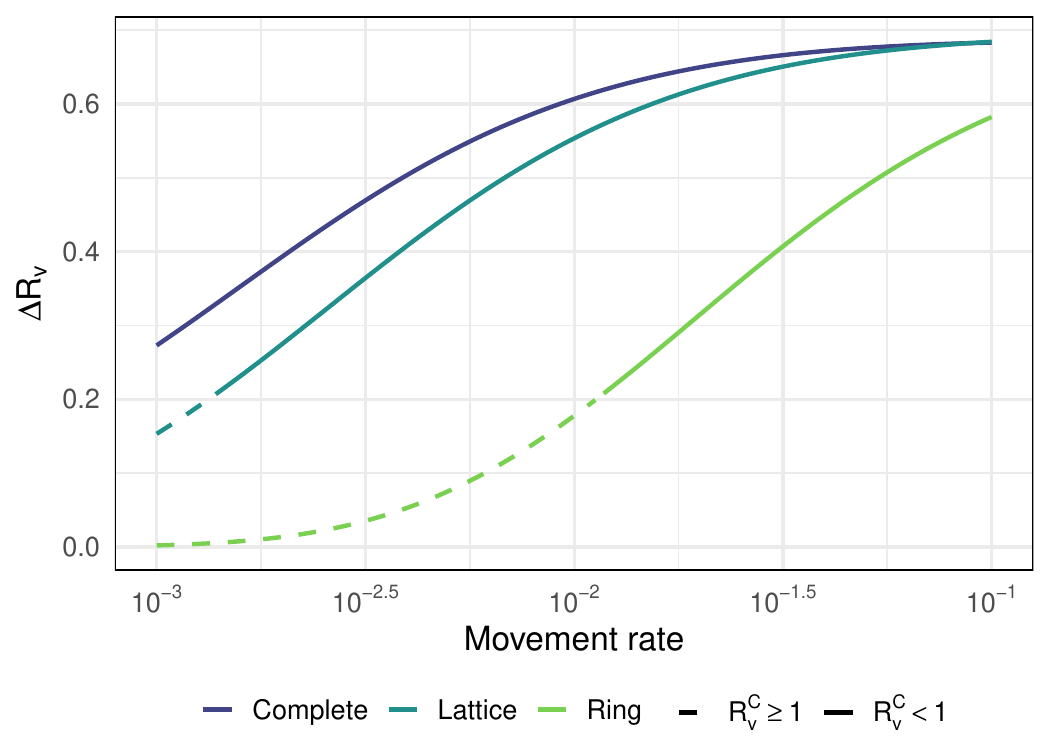}
        \caption{Single vaccinating patch}
        \label{fig:topology-rv-diff-single}
    \end{subfigure}
    \hfill
    \begin{subfigure}[b]{0.49\textwidth}
        \centering
        \includegraphics[width=\linewidth]{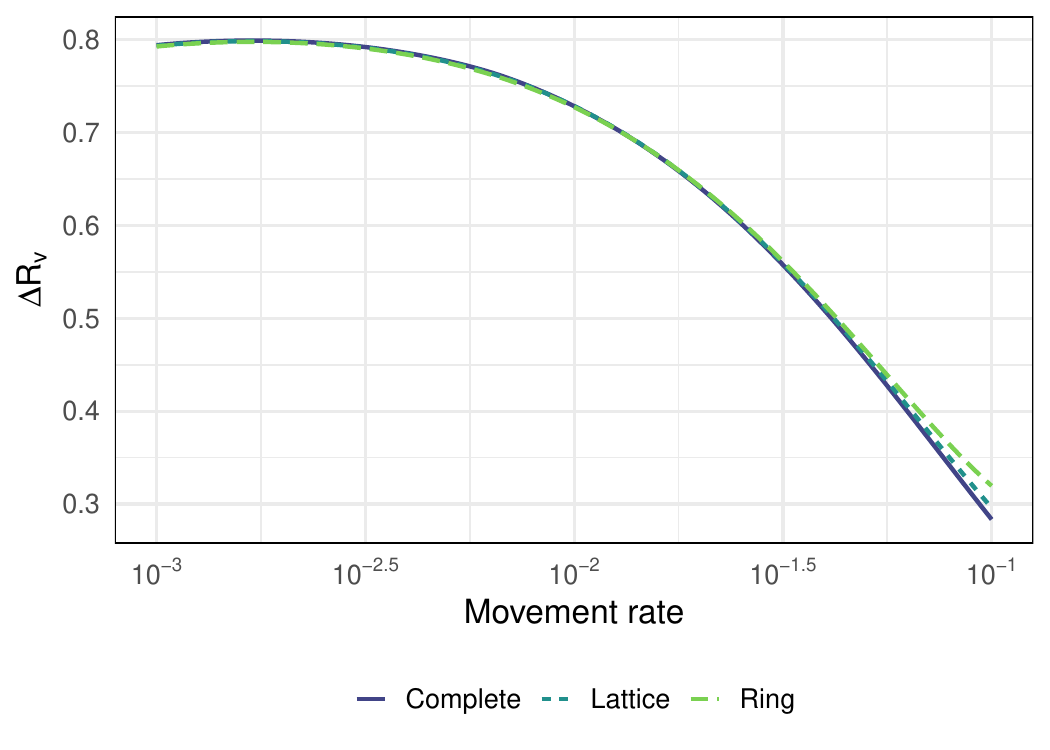}
        \caption{Vaccination of surrounding locations (core refuses)}
        \label{fig:topology-rv-diff-all-but-1}
    \end{subfigure}
    \caption{Difference $\Delta \mathcal{R}_v = \mathcal{R}_v^{\eqref{sys:nnSVIRS}} - \mathcal{R}_v^{\eqref{sys:nSVIRS}}$ in the system-wide vaccination reproduction number as a function of movement rate across three digraph topologies ($\phi = 0.01\text{ day}^{-1}$).
    (a) scenario where only a single patch vaccinates, (b) scenario where all but one patch vaccinate.}
    \label{fig:topology-rv-diff}
\end{figure*}

To illustrate how the difference between the two models depends on the movement network, the difference in the system-wide vaccination reproduction number $\Delta \mathcal{R}_v = \mathcal{R}_v^{\eqref{sys:nnSVIRS}} - \mathcal{R}_v^{\eqref{sys:nSVIRS}}$ is evaluated as a function of the movement rate $m$ at $\phi = 0.01\text{ day}^{-1}$.
Figure~\ref{fig:topology-rv-diff} compares $\Delta \mathcal{R}_v$ across the three regular digraph topologies: complete, lattice and ring.

The left panel displays a scenario where only a single patch vaccinates (Section~\ref{subsec:vacc_1patch}), while the right panel displays a scenario where a single core patch refuses to vaccinate (as in Section~\ref{subsec:vacc-allbut1patch}).

In the case where all but one patch vaccinate (Figure~\ref{fig:topology-rv-diff-all-but-1}), all three digraph types exhibit very similar $\Delta\R_v$ curves across $m$, showing that the influence of graph topology is minimal when 48 of 49 patches vaccinate and the main driver of $\Delta\R_v$ is the movement rate.

On the other hand, for the single vaccinating patch scenario (Figure~\ref{fig:topology-rv-diff-single}), the density of the digraph has a strong influence on the difference between reproduction numbers.
First, note that in all cases shown, $\R_v^{\eqref{sys:nnSVIRS}}>1$, whereas $\R_v^{\eqref{sys:nSVIRS}}$ is smaller than 1 except for the dashed part of the curves at lower movement rates.

In complete graphs (representing direct connections between all patches), $\mathcal{R}_v^{\eqref{sys:nSVIRS}}$ drops well below $\mathcal{R}_v^{\eqref{sys:nnSVIRS}}$ even at small movement rates, because visitors from every patch travel directly to the vaccinating patch.
Conversely, ring graphs (representing nearest-neighbour movement) preserve spatial heterogeneity up to much higher movement rates before the classic model homogenises vaccine coverage across the network.

\subsection{An example inspired by the 2003--2004 West and Central African polio crisis}
\label{sec:real_world}
The motivating example of the vaccination crisis that took place in 2003 and 2004 in northern Nigeria and led to the temporary suspension of oral polio vaccination is now considered.
The intent here is to illustrate how such a study can be conducted; parametrisation of this particular problem is indeed very complicated and this is more a thought experiment than a proper epidemiological study.
This example also serves to illustrate how different the behaviour is when classic and resident-traveller metapopulation models are used with more realistic graph architectures.

Movement data is scarce in general and even more so in the region under consideration. 
Except for air travel, most transportation in Africa is handled by \emph{informal transport} or \emph{paratransit}, i.e., operations run by micro-companies or individuals \cite{gil1989social,kerzhner2022informal}.
Also, trade between countries in the area is extremely active but also very poorly documented: it is estimated that ``informal cross border trade'', while escaping trade duties, amounts to upward of 40\% of the GDP and 55\% of employment in Africa \cite{gaarder2021towards,koroma2017formalization,larsen2025actor}.
For this reason, the volume of transportation happening between countries in the area affected by the 2003--2004 polio crisis is nigh impossible to estimate, although it can be assumed to be high.
The following scenario thus uses assumptions on the network and vaccination uptake.

\begin{figure}[htbp]
    \centering
    \includegraphics[width=\linewidth]{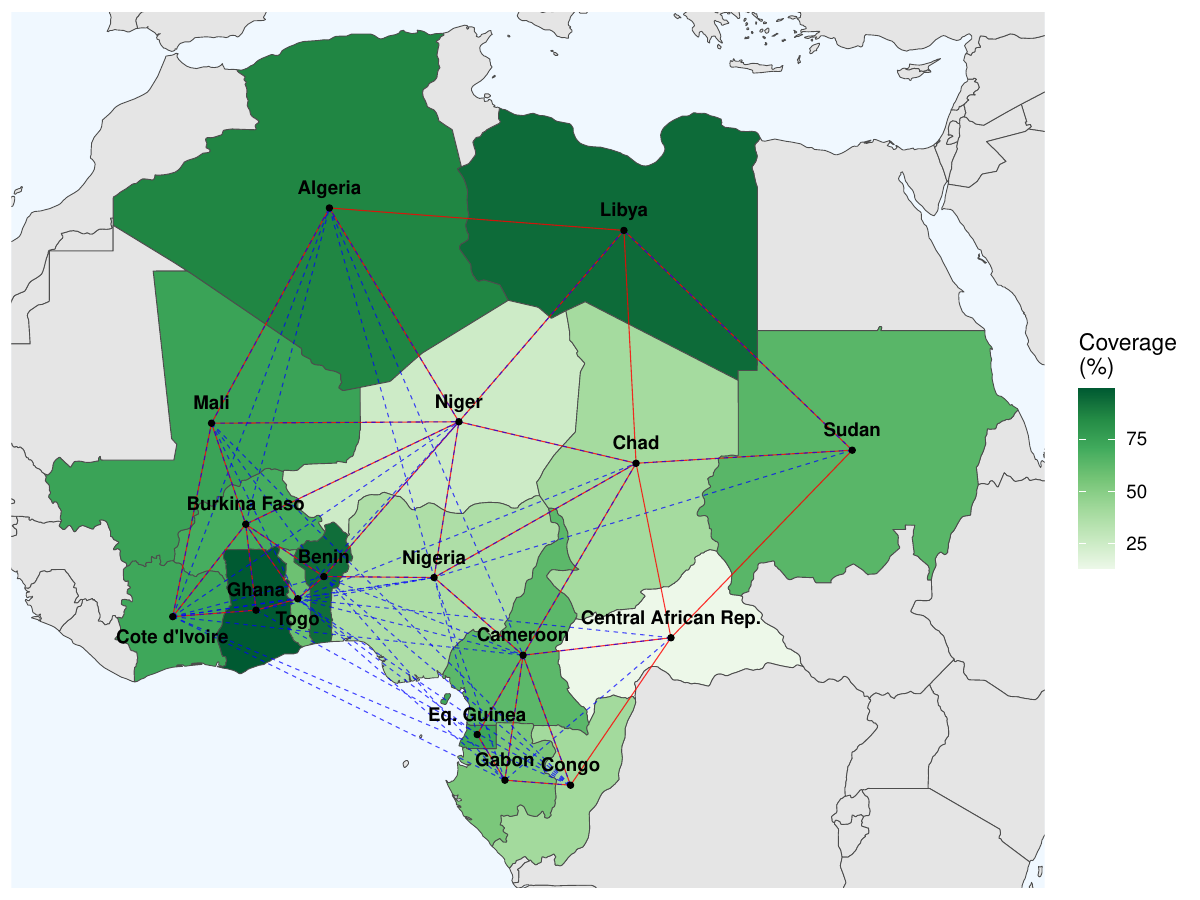}
    \caption{Percentage of children covered with the full three-doses regimen of oral polio vaccine in 2002 in countries at most two land borders away from Nigeria. 
    Ghana and C\^ote d'Ivoire are also included because of their proximity to Nigeria.
    Because 2002--2003 data for Nigeria were missing in the series, its 2004 value (37\%) is plotted here.
    Coverage values across these plotted countries range from 13\% (Central African Republic) to 99\% (Ghana). 
    The red network shows the graph of shared land borders representing short-range ground movement, while the blue network represents long-range air travel connections.
    Note that the graph vertices are country centroids, not capitals as in the actual transport graph.}
    \label{fig:map-vacc-coverage-Africa}
\end{figure}

\begin{figure}[htbp]
    \centering
    \includegraphics[width=\linewidth]{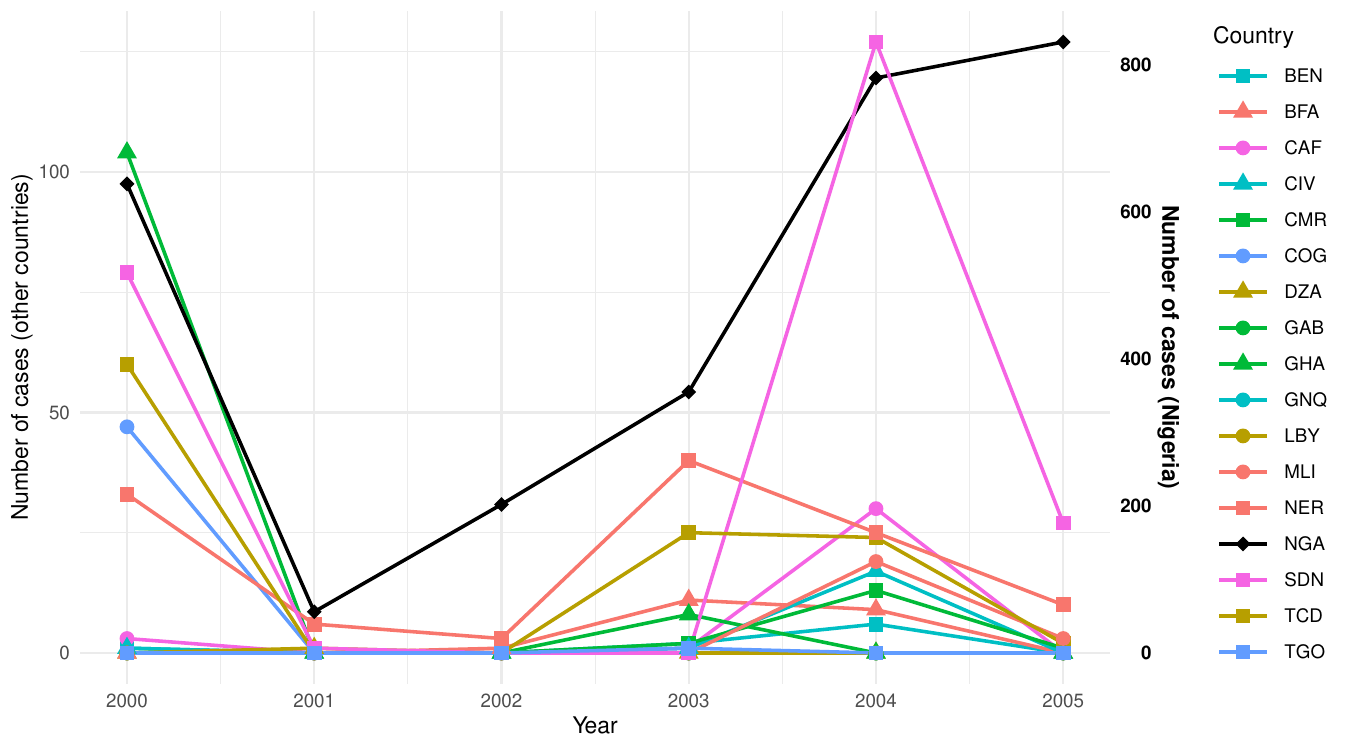}
    \caption{Reported cases of polio from 2000 to 2005 for the 17 countries in the region of interest. 
    Nigeria numbers are on the right axis. Note that the data consists of discrete annual points, which have been joined by lines for visualization.}
    \label{fig:polio-incidence-curves-Africa}
\end{figure}

Figure~\ref{fig:map-vacc-coverage-Africa} shows the percentage of children who had received the full three-doses OPV regimen in 2002 in Nigeria and countries at most two land borders away from Nigeria.
C\^ote d'Ivoire and Ghana were added to the list although they are three borders away because of their close proximity to Nigeria.
Specifically, Algeria, Benin, Burkina Faso, Cameroon, Central African Republic, Chad, Congo, C\^ote d'Ivoire, Equatorial Guinea, Gabon, Ghana, Libya, Mali, Niger, Nigeria, Sudan and Togo are considered.
The yearly reported incidence from 2000 to 2005 for these countries is shown in Figure~\ref{fig:polio-incidence-curves-Africa}.
For population and health data, United Nations, World Bank and World Health Organization sources are used.

As indicated, transportation data is scarce for this region, so the movement matrix is constructed as a combination of inferred short and long range travel matrices, both of which are derived from basic principles and hypotheses rather than actual data.
Short range ground travel is incorporated through the connection graph of shared land borders that has been derived, representing adjacent countries.
Long range air travel is considered through the global air transportation network, using the \texttt{wikipediaGATN} package \cite{arino_wikipediagatn}. 
For the air network, a base weight $A_{ji}$ from country $j$ to country $i$ is calculated as the sum of the number of airlines operating on all individual airport-to-airport routes connecting the two countries. 
For the ground network, the base weight $B_{ji}$ is simply 1 if $j$ and $i$ share a border, and 0 otherwise.

To compute the final connection rates for both networks, these base weights are multiplied by a gravity model that incorporates population sizes, distance and relative wealth. 
The standard gravity weight between countries $j$ and $i$ is taken as $N_j^{r\star} N_i^{r\star}/D_{ji}^2$, where $N_i^{r\star}$ denotes the population size of country $i$ and $D_{ji}$ is the great circle distance between their capital cities. 
To account for economic attractiveness, this gravity weight is multiplied by the destination's share of the combined GDP per capita of the two countries. 
Consequently, travel from a poorer origin to a wealthier destination forms a larger proportion of the bidirectional gravity weight. 
The final directional rate of movement from $j$ to $i$ for a given transport mode with base weight $W_{ji}$ ($A_{ji}$ for air or $B_{ji}$ for ground) is therefore computed as
\[
w_{ji}^{\text{mode}} = W_{ji} \ 
\frac{N_j^{r\star} N_i^{r\star}}{D_{ji}^2} \
\frac{\text{G}_i}{\text{G}_j + \text{G}_i}
\]
where $\text{G}$ stands for GDP per capita, and $\text{mode} \in \{\text{air}, \text{ground}\}$. 
The combined per-day rate of movement $w_{ji}$ from $j$ to $i$ is then defined as a linear combination of the air and ground connection rates,
\[
w_{ji} = \alpha_{\text{air}} w_{ji}^{\text{air}} + \alpha_{\text{ground}} w_{ji}^{\text{ground}},
\]
where $\alpha_{\text{air}}$ and $\alpha_{\text{ground}}$ are scaling parameters, taken as $2.26 \times 10^{-12}$ and $9.01 \times 10^{-12}$, respectively. These values were chosen to scale the combined rate of movement $w_{ji}$ such that its maximum value does not exceed $0.05\text{ day}^{-1}$. As a result, the non-zero values of $w_{ji}$ range from approximately $4.59 \times 10^{-7}$ to $0.0438\text{ day}^{-1}$.

In the resident-traveller model, residents of country $i$ leave their home patch for country $j$ at rate $m_{iji} = w_{ji}$ and return to their home patch at rate $m_{iij} = 1/14\text{ day}^{-1}$ (corresponding to an average trip duration of 14 days), with infectious individuals travelling at $10\%$ of the rate of other compartments ($\M^I = 0.1\M^S$).
To reflect socioeconomic and health-system heterogeneity across the 17 countries, the recovery period $1/\gamma_i$ is drawn uniformly in $[18, 24]$ days (consistent with the $\sim 20.4$-day mean duration of faecal poliovirus excretion in non-immunes \cite{FineCarneiro1999}), the isolated reproduction number $\R_{0i}$ is scaled linearly from $1.5$ (highest GDP per capita) to $3.5$ (lowest GDP per capita), and the relative susceptibility of vaccinated individuals $\pi_i$ is scaled linearly from $0.1$ (highest vaccine coverage) to $0.4$ (lowest vaccine coverage), with ongoing vaccination rates set to $\phi_i = \frac{c_i}{1-c_i}(\psi_i + d_i)$ (where $c_i$ is the country's reported coverage, capped at $0.99$) and initial conditions seeded with 10 infectious individuals in Nigeria.
Note that while estimates of the basic reproduction number of wild poliovirus in a completely naive population are higher, typically ranging from $6$ to $14$ in developing-country settings ($6$ to $11$ in medium-tier settings including northern Nigeria) \cite{FineCarneiro1999,DuintjerTebbens2013}, our simulation initialises all non-vaccinated under-5 children as fully susceptible ($R(0) = 0$) and does not explicitly model pre-existing natural or campaign-induced immunity from prior endemic circulation and supplementary immunisation activities.
The range $\R_{0i} \in [1.5, 3.5]$ therefore acts as an effective reproduction number under which well-vaccinated countries sit near or below the local control threshold while under-vaccinated countries remain supercritical.
Taking $N_i^{r\star}$ to be the under-5 population of country $i$ in the data, the home birth rate $b_{ii} = d_i N_i^{r\star}$ (with $b_{ij}=0$ for $j\neq i$) and equilibrium distribution $\bN_i^\star = (\D_i-\M_i)^{-1}\bb_i$ satisfy $\sum_{j=1}^{17} N_{ij}^\star = N_i^{r\star}$, so that \eqref{eq:dNi} takes the form
\[
\frac{d}{dt}\bN_i = (\M_{i}-\D_i)\bN_i+\bb_i=(\M_{i}-\D_i)(\bN_i-\bN_i^\star),
\]
and simulations starting with $\bN_i(0)=\bN_i^\star$ see no variation of the total resident population at the demographic level.

For the classic metapopulation model, the movement matrix $\M_c$ and demographic steady state $\bN_c^\star$ are obtained by lumping the resident-traveller model by current location at demographic equilibrium, so that the population present in each country ($N_{c,p}^\star = N_p^{p\star} = \sum_{i=1}^{17} N_{ip}^\star$) and the daily flux of travellers between every pair of countries match those of the resident-traveller model (whereas applying the raw one-way gravity rates $w_{ji}$ without return trips would artificially redistribute populations toward wealthier destinations).
The juxtaposition of the two model outputs illustrates how the classic model tends to homogenise vaccination coverage globally due to movement, while the resident-traveller model preserves the regional vulnerabilities associated with low vaccine coverage in specific countries.
Unlike the generic simulations explored earlier, for the specific case of poliomyelitis, the rate of immunity loss $\psi_{ij}$ is set to zero in the numerics, reflecting the fact that immunity is life-long or long enough that immunity loss is not relevant over the time scales considered here.

\begin{figure*}[htbp]
    \centering
    \begin{subfigure}[b]{0.49\textwidth}
        \centering
        \includegraphics[width=\linewidth]{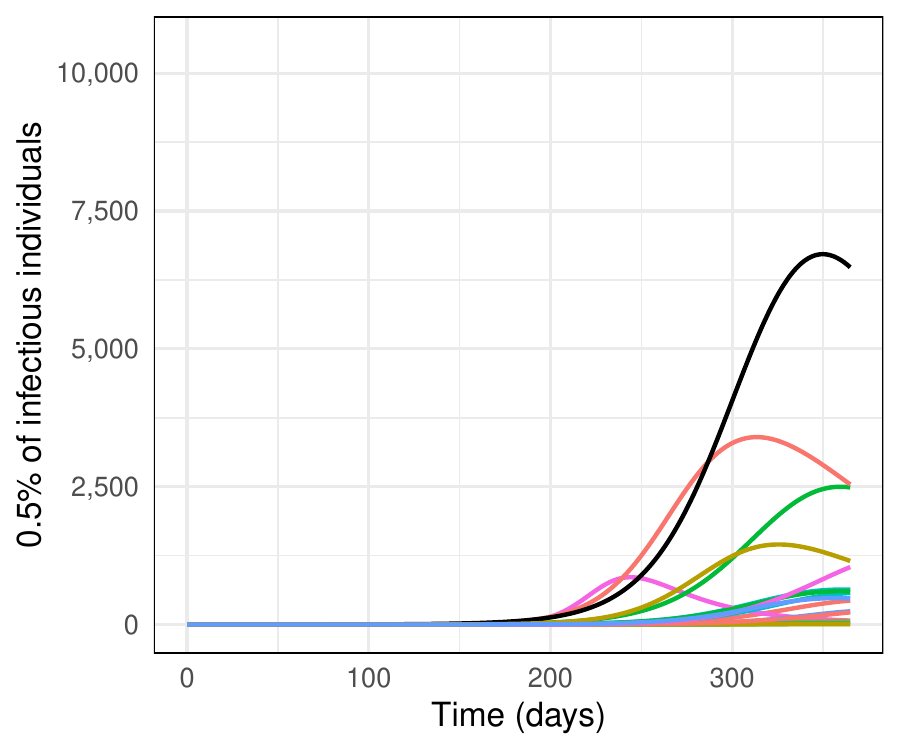}
        \caption{Classic metapopulation model}
        \label{fig:sims-WCA-curves-classic}
    \end{subfigure}
    \hfill
    \begin{subfigure}[b]{0.49\textwidth}
        \centering
        \includegraphics[width=\linewidth]{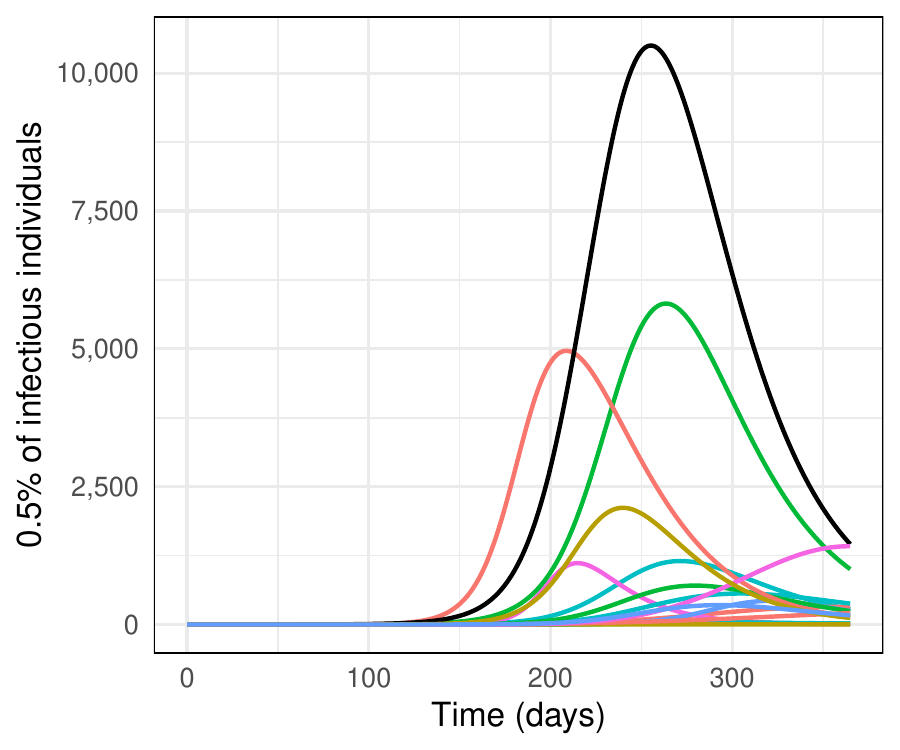}
        \caption{Resident-traveller model}
        \label{fig:sims-WCA-curves-rt}
    \end{subfigure}
    \caption{Infectious prevalence by current location, scaled by the paralysis-to-infection ratio ($0.5\%$ of $I_p^p(t)$) over a 365-day period on a common vertical scale, indicating the timing and relative size of the outbreak in each country. 
    Each curve represents a different country within the region (Nigeria in black). 
    (a) Classic metapopulation model; (b) resident-traveller metapopulation.}
    \label{fig:sims-WCA-curves}
\end{figure*}

\begin{figure*}[htbp]
    \centering
    \begin{subfigure}[b]{0.49\textwidth}
        \centering
        \includegraphics[width=\linewidth]{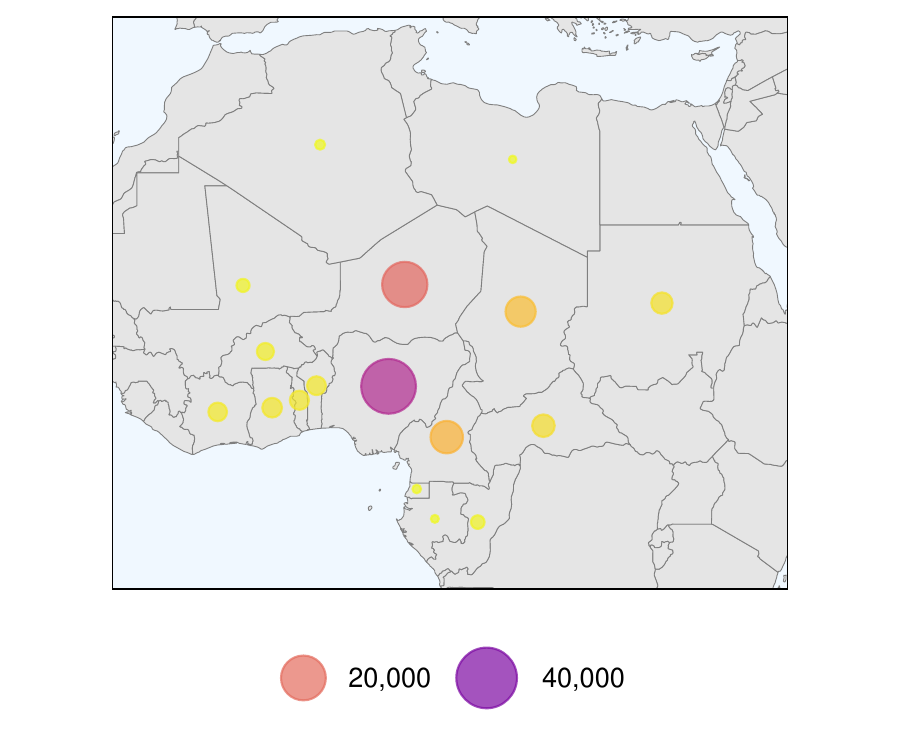}
        \caption{Classic metapopulation model}
        \label{fig:sims-WCA-maps-classic}
    \end{subfigure}
    \hfill
    \begin{subfigure}[b]{0.49\textwidth}
        \centering
        \includegraphics[width=\linewidth]{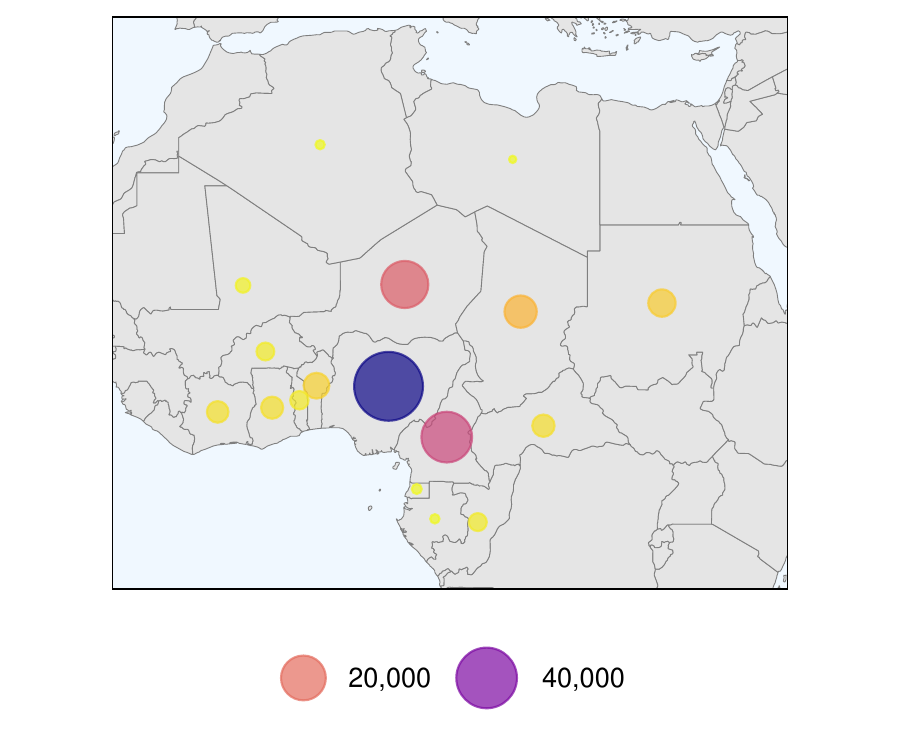}
        \caption{Resident-traveller model}
        \label{fig:sims-WCA-maps-rt}
    \end{subfigure}
    \caption{Geographic distribution of total accumulated paralytic cases after 1 year, tallied by country of infection and plotted on a shared size and colour scale. The classic model (a) predicts a smaller regional burden ($87{,}367$ paralytic cases overall, including $31{,}858$ in Nigeria and $9{,}158$ in Cameroon), whereas the resident-traveller model (b) predicts a substantially larger outbreak ($138{,}965$ paralytic cases overall, including $53{,}054$ acquired in Nigeria and $26{,}708$ in Cameroon).}
    \label{fig:sims-WCA-maps}
\end{figure*}

Figures~\ref{fig:sims-WCA-curves} and \ref{fig:sims-WCA-maps} present the simulated output of both models using this derived data, specifically focusing on the under-5 demographic.
Poliovirus infections are largely asymptomatic, with only roughly 0.5\% (1 in 200) of infections resulting in irreversible paralysis \cite{cdc_polio_disease,who_polio_factsheet}. 
To ensure an apples-to-apples comparison between the two formulations, all quantities in both figures are tallied by \emph{location of presence or infection} rather than by country of residence: Figure~\ref{fig:sims-WCA-curves} plots $0.5\%$ of the active infectious prevalence $I_p^p(t) = \sum_{i=1}^{17} I_{ip}(t)$ physically present in country $p$ over the course of a year, and Figure~\ref{fig:sims-WCA-maps} maps the accumulated paralytic cases ($0.5\%$ of cumulative new infections occurring inside country $p$, whether among local residents or visiting travellers) after one year, with both panels in each figure plotted on a common scale.
(Notice that in the resident-traveller model, cumulative cases can also be tallied by country of residence; because Nigeria's large under-5 cohort sends many unvaccinated travellers to neighbouring countries, $87{,}461$ paralytic cases occur among Nigerian residents overall, of which $53{,}054$ are acquired domestically within Nigeria and $34{,}407$ are acquired while visiting neighbouring countries such as Cameroon, Niger and Benin.)
While this thought experiment is not calibrated to reproduce the exact annual case counts of Figure~\ref{fig:polio-incidence-curves-Africa}, the resident-traveller formulation captures the qualitative pattern observed during the 2003--2004 resurgence: an earlier and larger epidemic wave in Nigeria accompanied by substantial transmission in neighbouring countries across the air and ground transportation networks ($138{,}965$ total paralytic cases region-wide after one year, compared with $87{,}367$ in the classic model, where the assimilation and vaccination of travellers in higher-coverage countries slows and dampens the regional epidemic).

Finally, two additional numerical sensitivity checks (not shown) are worth noting.
First, when larger values of the basic reproduction number typical of completely naive populations ($\R_{0i} \in [6, 11]$ or $[6, 14]$ \cite{FineCarneiro1999,DuintjerTebbens2013}) are used directly with $R(0) = 0$ and imperfect vaccine protection ($\pi_i \in [0.1, 0.4]$), there is much less difference between the outcomes of the classic and resident-traveller models: because even high vaccine coverage leaves an effective susceptible fraction $(1-c_i) + \pi_i c_i \ge \pi_i$, such large values of $\R_{0i}$ place every country well above the local epidemic threshold ($\R_{vi} \gg 1$) in both formulations, resulting in a rapid, synchronous epidemic wave across the entire region regardless of how traveller vaccination is bookkept.
Second, applying a 50\% global travel restriction, or a targeted 50\% reduction in travel solely for individuals leaving the outbreak epicenter (Nigeria), yields virtually no reduction in the total accumulated paralytic cases after one year in either model. 
Because the outbreak is driven heavily by local transmission dynamics once introduced to a patch, moderate travel restrictions delay the epidemic peak rather than substantially altering the cumulative one-year burden.

\section{Discussion}
\label{sec:discussion}
Comparing the resident-traveller framework with a classic Eulerian metapopulation model shows that tracking place of residence fundamentally alters predictions of vaccination impact across a mobility network.
Because classic metapopulation models assimilate travellers into the destination population, they apply local vaccination rates to visitors and export newly vaccinated individuals across the digraph.
In the high-mobility limit, this spatial mixing averages vaccine coverage globally, allowing high vaccination in one or more patches to drive the system-wide reproduction number $\R_v^{\eqref{sys:nSVIRS}}$ below unity even when other patches do not vaccinate at all.
In contrast, when vaccination is restricted to residents in their home patch, the resident-traveller reproduction number $\R_v^{\eqref{sys:nnSVIRS}}$ is bounded below by the digraph-adjusted basic reproduction number $\widetilde{\mathcal{R}}_0^k$ of any non-vaccinating patch $k$.
Consequently, whenever $\widetilde{\mathcal{R}}_0^k > 1$, unvaccinated subpopulations act as persistent reservoirs that prevent global elimination regardless of vaccination effort elsewhere.

From a mathematical standpoint, while Theorems~\ref{th:R0_nSIRS} and \ref{th:R0_nnSVIRS} establish the existence of at least one endemic equilibrium using uniform persistence theory when $\R_0>1$ (respectively $\R_v>1$), the problem of \emph{uniqueness} and \emph{global asymptotic stability} of the endemic equilibrium was not resolved here.
Furthermore, while Appendix~\ref{app:DFE_SIRS_GAS} proves the global asymptotic stability of the DFE for the baseline SIRS model under equal movement rates using a comparison theorem, establishing global stability for the DFE in the full SVIRS model when $\R_v^{\eqref{sys:nnSVIRS}} < 1$ remains open, as temporary loss-of-immunity dynamics and breakthrough infections allow the effective susceptible pool $S_{ij}+\pi_{ij}V_{ij}$ to transiently exceed its DFE value.
For resident-traveller models, this problem is notoriously complicated; the only work known to the authors addressing global stability in this context is \cite{IggidrSalletTsanou2012}, which proves a conjecture from \cite{ArinoVdD2003a} for the SIS case using the fact that the latter system is cooperative.
From numerical investigations, it is suspected that in the present model, this type of behaviour also prevails, with a unique endemic equilibrium existing and attracting all solutions when $\R_v>1$.

The model used here is intentionally simple in its description of epidemiological and demographic processes so as to isolate the structural role of residence-based vaccination, and several limitations should be noted when interpreting the numerical examples.
First, the model omits age structure, so demographic turnover in the under-5 poliomyelitis example is governed by the baseline mortality rate rather than explicit ageing out of the under-5 cohort, and routine childhood immunisation is approximated by a continuous rate $\phi_{ij}$.
Second, mixing within each patch is assumed to be homogeneous, and the numerical simulations assume that infectious individuals travel at $10\%$ of the rate of other compartments ($\M^I = 0.1\M^S$).
Third, the deterministic formulation does not capture stochastic fade-out at low prevalence or pathogen-specific features of oral poliovirus vaccines, such as vaccine-virus transmission or the emergence of circulating vaccine-derived polioviruses (cVDPV).
At the same time, the general formulation \eqref{sys:nnSVIRS} is flexible enough to accommodate non-resident vaccination policies through the off-diagonal rates $\phi_{ij} > 0$ ($i \neq j$), which can represent transit vaccination posts at border crossings or vaccination requirements for incoming travellers, as were eventually deployed during polio eradication campaigns.
It would also be interesting to undertake comparisons with residence-time (Lagrangian) models, which track individuals by home residence through a residence-time budgeting matrix rather than explicit movement fluxes (see, e.g., \cite{bichara2016vector,gao2024vector}).

This added realism comes at a computational cost, as the number of equations rises quickly with network size: for $n$ patches, a classic SVIRS metapopulation model requires $4n$ equations ($68$ for the $17$ countries in Section~\ref{sec:real_world}), whereas the resident-traveller formulation requires $4$ equations per active residence-location pair, up to $4n^2$ ($1{,}156$ for $n=17$) when all pairs of patches are connected, and requires specifying the zero/nonzero patterns of all movement matrices $\M_i$.

As a final remark, this paper was a long time coming: the authors started working on it in 2004.
For a variety of reasons, it was only finished recently.
While an immense global public health effort has culminated in the African continent being officially certified free of wild poliovirus in August 2020 \cite{Leke2020}, poliomyelitis is not yet eradicated globally.
Furthermore, vaccine coverage has decreased for other vaccine-preventable diseases, even in countries where vaccines are readily and widely accessible. 
Driven by rising vaccine hesitancy and disruptions from the COVID-19 pandemic, declining childhood immunization rates have recently fueled a severe resurgence of measles across Canada and the United States of America \cite{MathisRainesMastersEtAl2024,Rifai2024}, culminating in Canada officially losing its measles elimination status in November 2025 after more than 12 months of sustained transmission \cite{phac_measles_2025}.
With localised, under-vaccinated pockets allowing imported cases to spark explosive outbreaks, the core spatial mechanics of jurisdiction-based vaccination explored in this paper are arguably more relevant today than they were in 2004.

\subsection*{Acknowledgements}
The authors thank the Banff International Research Station (BIRS) for hosting the summer school and workshop during which this project was initiated.
JA acknowledges partial support from an NSERC Discovery Grant.
\appendix
\section{Global Asymptotic Stability of the DFE}
\label{app:DFE_SIRS_GAS}

\subsection{Proof for the SIRS model}
The global asymptotic stability (GAS) of the disease-free equilibrium (DFE) for the SIRS model \eqref{sys:nnSIRS} when $\mathcal{R}_0 < 1$ and movement rates are equal across all epidemiological states ($\M^S=\M^I=\M^R=\M$) is proved using the theory of asymptotically autonomous limit systems. 

From Lemma~\ref{lemma:behaviour_total_pop_epidemic_model}, it is known that under the assumption of equal movement rates for all epidemiological states, the total population in each subpopulation converges to its equilibrium, $\lim_{t\to\infty} N_{ij}(t) = N_{ij}^\star$. 
Because $S_{ij}(t) \le N_{ij}(t)$ for all $t$, it follows that for any $\eta > 0$, there exists a time $t_1 > 0$ such that for all $t > t_1$,
\[
S_{ij}(t) \le N_{ij}^\star + \eta \quad \text{and} \quad N_j^p(t) \ge N_j^{p\star} - \eta,
\]
(recalling that $N_j^p = \sum_k N_{kj}$ is the total population physically present in patch $j$).
The differential equations for the infectious compartments \eqref{sys:nnSIRS_dIij} can therefore be bounded above for $t > t_1$ by the linear differential inequality:
\[
\frac{d}{dt}\bI \le (F_\eta - W)\bI,
\]
where $W$ is defined as in \eqref{eq:W_matrix_SIRS}, and $F_\eta$ is the matrix of new infections evaluated using the perturbed upper bounds $N_{ij}^\star + \eta$ and lower bounds $N_j^{p\star} - \eta$. 

Because $F_\eta$ is a non-negative matrix and the off-diagonal entries of $-W = \M^I - \D - \Gamma$ consist of the non-negative movement rates from $\M^I$, the matrix $F_\eta - W$ has non-negative off-diagonal entries (i.e., it is a Metzler matrix). This property implies that the bounding linear differential system is cooperative. Thus, standard comparison principles for monotone flows (such as Kamke's comparison theorem) apply, despite the full nonlinear SIRS system not being cooperative \cite{Smith1995}.

When $\mathcal{R}_0 = \rho(FW^{-1}) < 1$, it follows (see, e.g., \cite[Theorem 2]{VdDWatmough2002}) that the spectral bound satisfies $s(F-W) < 0$. By the continuity of eigenvalues with respect to matrix entries, $\eta$ can be chosen sufficiently small such that $s(F_\eta - W) < 0$. 

By Kamke's comparison theorem for cooperative linear ODEs, $\bI(t)$ is bounded above by the solution to the linear system $\bX' = (F_\eta - W)\bX$. Since the spectral bound of $F_\eta - W$ is strictly negative, $\bX(t) \to \bm{0}$ as $t \to \infty$, which implies that $\lim_{t\to\infty} \bI(t) = \bm{0}$. 

Once $\bI \to \bm{0}$, the recovered equation \eqref{sys:nnSIRS_dRij} reduces asymptotically to $\frac{d}{dt}\bR = (\mathcal{M}^R - \D - \Psi)\bR$, which implies $\lim_{t\to\infty} \bR(t) = \bm{0}$ by Lemma~\ref{lemma:Rgoesto0_SIRS}. Finally, with both $\bI \to \bm{0}$ and $\bR \to \bm{0}$, the susceptible population must converge to the limiting total population, $\lim_{t\to\infty} \bS(t) = \bN^\star$. Thus, the DFE of the SIRS model is globally asymptotically stable.

\subsection{Adaptation to the SVIRS model}
It is natural to ask whether this bounding argument extends to the SVIRS model \eqref{sys:nnSVIRS} to prove global asymptotic stability of the DFE when $\mathcal{R}_v < 1$.

In the SIRS model, the DFE susceptible population is the entire population ($S_{ij}^\star = N_{ij}^\star$), meaning the physical bound $S_{ij}(t) \le N_{ij}(t)$ provides the tight asymptotic bound $S_{ij}^\star + \eta$. 
In the SVIRS model, the effective susceptible pool for subpopulation $(i,j)$ is $S_{ij}(t) + \pi_{ij}V_{ij}(t)$. Because $\pi_{ij} \le 1$, this pool is strictly bounded by the total living population:
\[
S_{ij}(t) + \pi_{ij}V_{ij}(t) \le S_{ij}(t) + V_{ij}(t) \le N_{ij}(t).
\]
Consequently, for any arbitrarily small $\eta > 0$ and sufficiently large $t$, $S_{ij}(t) + \pi_{ij}V_{ij}(t) \le N_{ij}^\star + \eta$. 
Substituting this upper bound into the infectious subsystem yields the same bounding matrix inequality $\bI' \le (F_\eta^0 - W)\bI$, where $F_\eta^0$ evaluates new infections assuming the entire population is fully susceptible. 
The spectral radius $\rho(F^0 W^{-1})$ corresponds to the baseline, \emph{vaccine-free} basic reproduction number $\R_{0}^{\eqref{sys:nnSIRS}}$. 
Thus, if $\R_{0}^{\eqref{sys:nnSIRS}} < 1$, Kamke's theorem proves that the SVIRS disease-free equilibrium is globally asymptotically stable. 

However, establishing global asymptotic stability specifically when $\R_v^{\eqref{sys:nnSVIRS}} < 1 < \R_{0}^{\eqref{sys:nnSIRS}}$ using this comparison method is not straightforward. 
The issue lies in bounding the effective transmission pool $S_{ij}(t) + \pi_{ij}V_{ij}(t)$ tightly around its DFE value $S_{ij}^\star + \pi_{ij}V_{ij}^\star$. 
Unlike in the baseline SIRS model, where the natural excursion of individuals through the infectious and recovered classes ($S \to I \to R \to S$) is strictly bounded by the demographic ceiling $N_{ij}(t)$ with $S_{ij}^\star = N_{ij}^\star$, temporary immunity ($\psi_{ij}>0$) and breakthrough infections ($V \to I \to R \to S$) in the SVIRS model can move individuals from the partially susceptible vaccinated class ($\pi_{ij}<1$) back into the fully susceptible class, allowing $S_{ij}(t) + \pi_{ij}V_{ij}(t)$ to transiently exceed $S_{ij}^\star + \pi_{ij}V_{ij}^\star$. The standard comparison theorem thus fails. 
Establishing Global Asymptotic Stability (GAS) for the SVIRS resident-traveller model, either at the DFE or for the endemic equilibrium, remains a non-trivial open problem. Standard graph-theoretic Lyapunov methods \cite{shuai2013global} cannot be readily applied due to the distinct topologies of mobility and infection, alongside the non-cooperative nature of standard incidence mixing.
\section{The classic metapopulation SVIRS model}
\label{app:classic-SVIRS}
Part of the justification for using a resident-traveller framework is that it allows consideration of problems such as those that arise when one location does not vaccinate while others do and residents maintain their vaccination status and behaviour even when travelling.
In order to illustrate this claim, a classic metapopulation model analogous to \eqref{sys:nnSVIRS} (with no vaccination at birth, matching $P_i=0$ in the numerical comparisons) is now considered.
In classic metapopulation models, individuals who travel from patch $j$ to patch $i$ immediately become, upon arrival in $i$, members of patch $i$. Consequently, they become subject to the vaccination policy and mixing patterns of their destination patch.

The classic metapopulation model takes the following form, where, for simplicity, it is assumed that movement is equal for all epidemiological states and $m_{ij}$ denotes the rate of movement from patch $j$ to patch $i$.
Let $N_i=S_i+I_i+R_i+V_i$ be the total population in patch $i$. The dynamics are governed by:
\begin{subequations}\label{sys:nSVIRS}
\begin{align}
\frac{d}{dt}S_{i} &= 
\birth_{i}+\psi_{i} R_{i} 
-\left(d_{i}+\phi_{i}
+\beta_{i}\frac{I_i}{N_i}\right)
S_{i}+\sum_{j=1}^n m_{ij} S_{j} 
\label{sys:nSVIRS_dSi} \\
\frac{d}{dt}I_{i} &= 
\left(S_{i}+\pi_{i} V_{i}\right)
\beta_{i}\frac{I_{i}}{N_i}
-\left(d_{i}+\gamma_{i}\right)I_{i}
+\sum_{j=1}^n m_{ij}I_{j}
\label{sys:nSVIRS_dIi} \\
\frac{d}{dt}R_{i} &= \gamma_{i}I_{i}
-\left(d_{i}+\psi_{i}\right)R_{i}
+\sum_{j=1}^n m_{ij}R_{j}
\label{sys:nSVIRS_dRi} \\
\frac{d}{dt}V_{i} &= \phi_{i}S_{i} 
-\left(d_{i}
+\pi_{i}\beta_{i}\frac{I_{i}}{N_i}
\right)V_{i}
+\sum_{j=1}^n m_{ij}V_{j}.
\label{sys:nSVIRS_dVi}
\end{align}
\end{subequations}

A basic analysis follows steps from standard metapopulation models \cite{Arino2009}.
The vector form of the system involves vectors $\bX=(X_1,\ldots,X_n)^T$ for $X\in\{N,S,V,I,R\}$, $\bb=(b_1,\ldots,b_n)^T$ and diagonal matrices $\D_c=\diag(d_1,\ldots,d_n)$, $\Phi_c=\diag(\phi_1,\ldots,\phi_n)$, $\Gamma_c=\diag(\gamma_1,\ldots,\gamma_n)$ and $\Psi_c=\diag(\psi_1,\ldots,\psi_n)$.
Let $\M_c$ be the movement matrix with off-diagonal entries $m_{ij}$ and diagonal entries $m_{ii}=-\sum_{k\neq i} m_{ki}$.
Because movement is equal for all epidemiological states, the total population satisfies
\[
\frac{d}{dt}\bN = \bb-\D_c\bN+\M_c\bN,
\]
so that $\lim_{t\to\infty}\bN(t)=\bN^\star_c:=(\D_c-\M_c)^{-1}\bb$. 
At the disease-free equilibrium (DFE), there are no infectious or recovered individuals ($\bI_c^\star=\bR_c^\star=0$), and 
\begin{align*}
0 &= \bb-(\D_c+\Phi_c-\M_c)\bS \\
0 &= \Phi_c\bS - (\D_c-\M_c)\bV
\end{align*}
which yields the unique positive equilibrium
\begin{equation}\label{eq:DFE_classic}
\bS_c^\star=(\Phi_c+\D_c-\M_c)^{-1}\bb, \qquad \bV_c^\star=(\D_c-\M_c)^{-1}\Phi_c\bS_c^\star.
\end{equation}

The local asymptotic stability of the DFE is governed by the vaccination reproduction number $\R_v^{\eqref{sys:nSVIRS}}=\rho(F_c W_c^{-1})$, determined by the next-generation matrix method \cite{VdDWatmough2002}.

\begin{theorem}
\label{th:R0_classic_SVIRS}
The DFE of the classic metapopulation model is locally asymptotically stable if $\R_v^{\eqref{sys:nSVIRS}} < 1$ and unstable if $\R_v^{\eqref{sys:nSVIRS}} > 1$.
\end{theorem}

In this setup, the rate of new infections in patch $i$ is $\mathcal{F}_i = (S_i+\pi_i V_i)\beta_i \frac{I_i}{N_i}$, and the other transitions are given by $\mathcal{V}_i = (d_i+\gamma_i)I_i - \sum_{j=1}^n m_{ij} I_j$.
Linearizing at the DFE, the non-negative matrix $F_c$ of new infections and the non-singular M-matrix $W_c$ are
\[
F_c = \diag\left( \beta_i \frac{S_{ci}^\star + \pi_i V_{ci}^\star}{N_{ci}^\star} \right) \quad \text{and} \quad W_c = \Gamma_c + \D_c - \M_c.
\]
Similarly, the basic reproduction number $\R_0^\eqref{sys:nSVIRS}$ in the absence of vaccination ($\Phi_c=0$, giving $\bV_c^\star=0$ and $\bS_c^\star=\bN_c^\star$) is $\R_0^\eqref{sys:nSVIRS} = \rho(F_c^{0} W_c^{-1})$, where $F_c^{0} = \diag(\beta_i)$.

\bibliographystyle{plain}
\bibliography{references}

@article{bichara2016vector,
  title={Vector-borne diseases models with residence times--a {L}agrangian perspective},
  author={Bichara, D. and Castillo-Chavez, C.},
  journal={Mathematical Biosciences},
  volume={281},
  pages={128--138},
  year={2016},
  publisher={Elsevier}
}

@article{gao2024vector,
  title={Vector-borne disease models with {L}agrangian approach},
  author={Gao, D. and Cao, L.},
  journal={Journal of Mathematical Biology},
  volume={88},
  number={2},
  pages={22},
  year={2024},
  publisher={Springer}
}

@techreport{koroma2017formalization,
	title={Formalization of informal trade in {A}frica: {T}rends, experiences and socio-economic impact},
	author={Koroma, S. and Nimarkoh, J. and You, N. and Ogalo, V. and Owino, B.},
	year={2017},
	institution={Food and Agriculture Organization of the United Nations}
}

@techreport{gil1989social,
	title={Social and labour aspects of urban passenger transport in selected {A}frican cities},
	author={Gil, A.},
	year={1989},
	institution={International Labour Organisation}
}

@techreport{gaarder2021towards,
	title={Towards an estimate of informal cross-border trade in {A}frica},
	author={Gaarder, E. and Luke, D. and Sommer, L.},
	year={2021},
	institution={United Nations Economic Commission for Africa}
}

@article{larsen2025actor,
	title={An actor-based perspective on informal cross-border trading},
	author={Larsen, M.M. and Namatovu, R. and Narula, R.},
	journal={The Journal of Development Studies},
	volume={62},
	number={8},
	pages={1275--1295},
	year={2026},
	doi={10.1080/00220388.2025.2569391},
	publisher={Taylor \& Francis}
}

@article{kerzhner2022informal,
	title={Is informal transport flexible?},
	author={Kerzhner, T.},
	journal={Journal of Transport and Land Use},
	volume={15},
	number={1},
	pages={671--689},
	year={2022},
	publisher={JSTOR}
}

@article{shuai2013global,
	title={Global stability of infectious disease models using {L}yapunov functions},
	author={Shuai, Z. and van den Driessche, P.},
	journal={SIAM Journal on Applied Mathematics},
	volume={73},
	number={4},
	pages={1513--1532},
	year={2013},
	publisher={SIAM}
}

@article{Khan2002,
  author  = {Khan, K. and Muennig, P. and Behta, M. and Zivin, J. G.},
  title   = {Global drug-resistance patterns and the management of latent tuberculosis infection in immigrants to the {U}nited {S}tates},
  journal = {New England Journal of Medicine},
  volume  = {347},
  number  = {23},
  pages   = {1850--1859},
  year    = {2002},
  doi     = {10.1056/NEJMsa021099},
  pmid    = {12466510}
}

@article{Schwartzman2005,
  author  = {Schwartzman, K. and Oxlade, O. and Barr, R. G. and Grimard, F. and Acosta, I. and Baez, J. and Ferreira, E. and Melgen, R. E. and Morose, W. and Salgado, A. C. and Jacquet, V. and Maloney, S. and Laserson, K. and Mendez, A. P. and Menzies, D.},
  title   = {Domestic returns from investment in the control of tuberculosis in other countries},
  journal = {New England Journal of Medicine},
  volume  = {353},
  number  = {10},
  pages   = {1008--1020},
  year    = {2005},
  doi     = {10.1056/NEJMsa043194},
  pmid    = {16148286}
}

@article{Leke2020,
  author  = {Fomban Leke, R. G. and King, A. and Pallansch, M. A. and Tangermann, R. H. and Mkanda, P. and Chunsuttiwat, S. and Jack, A. and Kabor\'e, B. J. and Kane, I. and Khomo, N. E. and Lopes-Feio, R. J. and Maiga, Z. M. and Matuja, W. B. P. and Kironde, N. E. and Kouassi, B. and Sadrizadeh, B. and Tomori, O.},
  title   = {Certifying the interruption of wild poliovirus transmission in the {WHO} {A}frican region on the turbulent journey to a polio-free world},
  journal = {The Lancet Global Health},
  volume  = {8},
  number  = {10},
  pages   = {e1345--e1351},
  year    = {2020},
  doi     = {10.1016/S2214-109X(20)30382-X}
}

@article{Rifai2024,
  author  = {Rifai, A. A. and Pittman Ratterree, D. C. and Kong, J. D. and Ndeffo-Mbah, M. L.},
  journal = {PLOS Global Public Health},
  title   = {Epidemiological review on the resurgence of measles outbreaks in {C}anada during the post-elimination era: {A} scoping review},
  volume  = {6},
  number  = {4},
  pages   = {e0006295},
  year    = {2026},
  doi     = {10.1371/journal.pgph.0006295}
}

@article{MathisRainesMastersEtAl2024,
  author  = {Mathis, A.D. and Raines, K. and Masters, N.B. and Filardo, T.D. and Kim, G. and Crooke, S.N. and Bankamp, B. and Rota, P.A. and Sugerman, D.E.},
  journal = {MMWR Morbidity and Mortality Weekly Report},
  pages   = {295--300},
  title   = {Measles --- {U}nited {S}tates, {J}anuary 1, 2020--{M}arch 28, 2024},
  volume  = {73},
  number  = {14},
  year    = {2024},
  doi     = {10.15585/mmwr.mm7314a1}
}

@article{Ghinai2013,
  author  = {Ghinai, I. and Willott, C. and Dadari, I. and Larson, H.J.},
  journal = {Global Public Health},
  number  = {10},
  pages   = {1138--1150},
  title   = {Listening to the rumours: what the northern {N}igeria polio vaccine boycott can tell us ten years on},
  volume  = {8},
  year    = {2013},
  doi     = {10.1080/17441692.2013.859720}
}

@article{ChristianPoutanenLouftyEtAl2004,
  author   = {Christian, M.D. and Poutanen, S.M. and Loutfy, M.R. and
              Muller, M.P. and Low, D.E.},
  journal  = {Clinical Infectious Diseases},
  month    = may,
  pages    = {1420--1427},
  title    = {Severe Acute Respiratory Syndrome},
  volume   = {38},
  year     = {2004},
  doi      = {10.1086/420743},
  issn     = {1537-6591}
}

@article{RuanWangLevin2006,
  author  = {Ruan, S. and Wang, W. and Levin, S.A.},
  journal = {Mathematical Biosciences and Engineering},
  number  = {1},
  pages   = {205-218},
  title   = {The effect of global travel on the spread of {SARS}},
  volume  = {3},
  year    = {2006}
}

@article{KhanArinoHuEtAl2009,
  author    = {Khan, K. and Arino, J. and Hu, W. and Raposo, P. and
               Sears, J. and Calderon, F. and Heidebrecht, C. and
               Macdonald, M. and Liauw, J. and Chan, A. and
               Gardam, M.},
  journal   = {New England Journal of Medicine},
  number    = {2},
  pages     = {212--214},
  title     = {Spread of a novel influenza {A} ({H1N1}) virus via
               global airline transportation},
  volume    = {361},
  year      = {2009},
  doi       = {10.1056/NEJMc0904559},
  url       = {http://www.nejm.org/doi/full/10.1056/NEJMc0904559}
}

@article{Arino2022,
  author   = {Arino, J.},
  journal  = {Fields Institute Communications},
  pages    = {25-51},
  title    = {Describing, modelling and forecasting the spatial and
              temporal spread of {COVID-19} -- A short review},
  volume   = {85},
  year     = {2022},
  doi      = {10.1007/978-3-030-85053-1_2}
}

@article{KhanMemishChabbraEtAl2010,
  author    = {Khan, K. and Memish, Z.A. and Chabbra, A. and
               Liauw, J. and Hu, W. and Janes, D.A. and Sears, J. and
               Arino, J. and Macdonald, M. and Calderon, F. and
               Raposo, P. and Heidebrecht, C. and Wang, J. and
               Chan, A. and Brownstein, J. and Gardam, M.},
  journal   = {Journal of Travel Medicine},
  number    = {2},
  pages     = {75--81},
  title     = {Global public health implications of a mass gathering
               in {M}ecca, {S}audi {A}rabia during the midst of an
               influenza pandemic},
  volume    = {17},
  year      = {2010},
  doi       = {10.1111/j.1708-8305.2010.00397.x},
  url       = {Khan_etal-2010-JTM17.pdf}
}

@article{BereraZambon2013,
  author      = {Berera, D. and Zambon, M.},
  institution = {Division of International Epidemiology and Population
                 Studies, Fogarty International Center, National
                 Institutes of Health, Bethesda, MD, USA; College of
                 Medicine, University of Central Florida, Orlando, FL,
                 USA.},
  journal     = {Influenza Other Respiratory Viruses},
  month       = {Nov},
  pages       = {72--79},
  title       = {Antivirals in the 2009 pandemic--lessons and
                 implications for future strategies.},
  volume      = {7 Suppl 3},
  year        = {2013},
  doi         = {10.1111/irv.12172},
  language    = {eng},
  url         = {http://dx.doi.org/10.1111/irv.12172}
}

@article{LangloisklassenWooldrageManfredaEtAl2011,
  author      = {Langlois-Klassen, D. and Wooldrage, K.M. and
                 Manfreda, J. and Sutherland, K. and Ellis, E. and
                 Phypers, M. and Gushulak, B. and Long, R.},
  institution = {Department of Public Health Sciences, University of
                 Alberta, Edmonton, AB, Canada.},
  journal     = {Eur Respir J},
  month       = {Oct},
  number      = {4},
  pages       = {895--902},
  title       = {Piecing the puzzle together: foreign-born
                 tuberculosis in an immigrant-receiving country.},
  volume      = {38},
  year        = {2011},
  doi         = {10.1183/09031936.00196610},
  language    = {eng},
  url         = {http://dx.doi.org/10.1183/09031936.00196610}
}

@article{ClementsGreenoughShull2006,
  author      = {Clements, C.J. and Greenough, P. and Shull, D.},
  institution = {Centre for International Health, The Macfarlane
                 Burnet Institute for Medical Research and Public
                 Health Ltd., Melbourne, Australia. john@clem.com.au},
  journal     = {Curr Drug Saf},
  month       = {Jan},
  number      = {1},
  pages       = {117--119},
  title       = {How vaccine safety can become political--the example
                 of polio in {N}igeria},
  volume      = {1},
  year        = {2006},
  language    = {eng}
}

@article{Jegede2007,
  author      = {Jegede, A.S.},
  institution = {University of Toronto Joint Centre for Bioethics,
                 Toronto, ON, Canada. sayjegede@yahoo.com},
  journal     = {PLoS Med},
  month       = {Mar},
  number      = {3},
  pages       = {e73},
  title       = {What led to the {N}igerian boycott of the polio
                 vaccination campaign?},
  volume      = {4},
  year        = {2007},
  doi         = {10.1371/journal.pmed.0040073},
  language    = {eng},
  url         = {http://dx.doi.org/10.1371/journal.pmed.0040073}
}

@article{PallanschSandhu2006,
  author      = {Pallansch, M.A. and Sandhu, H.S.},
  institution = {National Center for Immunization and Respiratory
                 Diseases, Centers for Disease Control and Prevention,
                 Atlanta, USA.},
  journal     = {N Engl J Med},
  month       = {Dec},
  number      = {24},
  pages       = {2508--2511},
  title       = {The eradication of polio--progress and challenges.},
  volume      = {355},
  year        = {2006},
  doi         = {10.1056/NEJMp068200},
  language    = {eng},
  url         = {http://dx.doi.org/10.1056/NEJMp068200}
}

@article{Fleck2004a,
  author    = {Fleck, F.},
  journal   = {BMJ},
  month     = {Jul},
  number    = {7457},
  pages     = {70},
  title     = {Nigerian state of {K}ano resumes polio vaccination},
  volume    = {329},
  year      = {2004},
  doi       = {10.1136/bmj.329.7457.70-d},
  language  = {eng},
  url       = {http://dx.doi.org/10.1136/bmj.329.7457.70-d}
}

@article{KaufmannFeldbaum2009,
  author      = {Kaufmann, J.R. and Feldbaum, H.},
  institution = {Global Health and Foreign Policy Initiative, Paul H.
                 Nitze School of Advanced International Studies, in
                 Arlington, Virginia, USA. kaufmannjr2@aol.com},
  journal     = {Health Affairs},
  number      = {4},
  pages       = {1091--1101},
  title       = {Diplomacy and the polio immunization boycott in
                 {N}orthern {N}igeria.},
  volume      = {28},
  year        = {2009},
  doi         = {10.1377/hlthaff.28.4.1091},
  language    = {eng},
  url         = {http://dx.doi.org/10.1377/hlthaff.28.4.1091}
}

@article{KlepacMetcalfMcLeanHampson2013,
  author      = {Klepac, P. and Metcalf, C.J.E. and McLean, A.R. and
                 Hampson, K.},
  institution = {Department of Applied Mathematics and Theoretical
                 Physics, University of Cambridge, Wilberforce Road,
                 Cambridge CB3 0WA, UK.},
  journal     = {Philos Trans R Soc Lond B Biol Sci},
  month       = {Aug},
  number      = {1623},
  pages       = {20120137},
  title       = {Towards the endgame and beyond: complexities and
                 challenges for the elimination of infectious
                 diseases.},
  volume      = {368},
  year        = {2013},
  doi         = {10.1098/rstb.2012.0137},
  language    = {eng},
  url         = {http://dx.doi.org/10.1098/rstb.2012.0137}
}

@article{KnolUrbanusE.M.MollemaEtAl2013,
  author    = {Knol, M.J. and Urbanus, A.T. and Swart, E.M. and
               Mollema, L. and Ruijs, W.L. and van Binnendijk, R.S. and
               te Wierik, M.J. and de Melker, H.E. and Timen, A. and
               Hahn\'e, S.J.},
  journal   = {Eurosurveillance},
  number    = {36},
  pages     = {2-7},
  title     = {Large ongoing measles outbreak in a religious
               community in the {N}etherlands since {M}ay 2013.},
  volume    = {18},
  year      = {2013}
}

@article{DownsBruinFischhoff2008,
  author      = {Downs, J.S. and Bruine de Bruin, W. and
                 Fischhoff, B.},
  institution = {of Social and Decision Sciences, Pittsburgh, PA
                 15213-3890, United States. downs@cmu.edu},
  journal     = {Vaccine},
  month       = {Mar},
  number      = {12},
  pages       = {1595--1607},
  title       = {Parents' vaccination comprehension and decisions.},
  volume      = {26},
  year        = {2008},
  doi         = {10.1016/j.vaccine.2008.01.011},
  language    = {eng},
  url         = {http://dx.doi.org/10.1016/j.vaccine.2008.01.011}
}

@article{Grabenstein2013,
  author      = {Grabenstein, J.D.},
  institution = {Merck Vaccines, 770 Sumneytown Pike, WP97-B364, West
                 Point, PA 19426, USA. john_grabenstein@merck.com},
  journal     = {Vaccine},
  month       = {Apr},
  number      = {16},
  pages       = {2011--2023},
  title       = {What the world's religions teach, applied to vaccines
                 and immune globulins.},
  volume      = {31},
  year        = {2013},
  doi         = {10.1016/j.vaccine.2013.02.026},
  language    = {eng},
  url         = {http://dx.doi.org/10.1016/j.vaccine.2013.02.026}
}

@article{WeigelWeitmannRautmannSchmidtEtAl2014,
  author      = {Weigel, M. and Weitmann, K. and Rautmann, C. and
                 Schmidt, J. and Bruns, R. and Hoffmann, W.},
  institution = {1 Department Epidemiology of Health Care and
                 Community Health, Institute for Community Medicine,
                 University Medicine Greifswald, Greifswald, Germany.},
  journal     = {Eur J Public Health},
  volume      = {24},
  number      = {6},
  pages       = {1009--1016},
  title       = {Impact of physicians' attitude to vaccination on
                 local vaccination coverage for pertussis and measles
                 in {G}ermany.},
  year        = {2014},
  doi         = {10.1093/eurpub/cku013},
  language    = {eng},
  url         = {http://dx.doi.org/10.1093/eurpub/cku013}
}

@article{IsaacsKilhamLeaskTobin2009,
  author    = {Isaacs, D. and Kilham, H. and Leask, J. and
               Tobin, B.},
  journal   = {Vaccine},
  month     = {Jan},
  number    = {5},
  pages     = {615--618},
  title     = {Ethical issues in immunisation.},
  volume    = {27},
  year      = {2009},
  doi       = {10.1016/j.vaccine.2008.11.002},
  language  = {eng},
  url       = {http://dx.doi.org/10.1016/j.vaccine.2008.11.002}
}

@article{BedfordElliman2000,
  author      = {Bedford, H. and Elliman, D.},
  institution = {Department of Epidemiology, Institute of Child
                 Health, London WC1N 1EH.},
  journal     = {BMJ},
  month       = {Jan},
  number      = {7229},
  pages       = {240--243},
  title       = {Concerns about immunisation.},
  volume      = {320},
  year        = {2000},
  language    = {eng}
}

@article{Blume2006,
  author      = {Blume, S.},
  institution = {Department of Sociology and Anthropology, University
                 of Amsterdam, o.z. Achterburgwal 185, 1012 DK
                 Amsterdam, The Netherlands. s.s.blume@uva.nl},
  journal     = {Soc Sci Med},
  month       = {Feb},
  number      = {3},
  pages       = {628--642},
  title       = {Anti-vaccination movements and their
                 interpretations.},
  volume      = {62},
  year        = {2006},
  doi         = {10.1016/j.socscimed.2005.06.020},
  language    = {eng},
  url         = {http://dx.doi.org/10.1016/j.socscimed.2005.06.020}
}

@article{Kata2010,
  author      = {Kata, A.},
  institution = {Department of Anthropology, Chester New Hall,
                 McMaster University, 1280 Main St. W, Hamilton,
                 Ontario L8S 4L8, Canada. aniakata@gmail.com},
  journal     = {Vaccine},
  month       = {Feb},
  number      = {7},
  pages       = {1709--1716},
  title       = {A postmodern {P}andora's box: anti-vaccination
                 misinformation on the {I}nternet.},
  volume      = {28},
  year        = {2010},
  doi         = {10.1016/j.vaccine.2009.12.022},
  language    = {eng},
  url         = {http://dx.doi.org/10.1016/j.vaccine.2009.12.022}
}

@article{Kata2012,
  author      = {Kata, A.},
  institution = {McMaster University, Psychiatry and Behavioural
                 Neurosciences, 555 Sanatorium Road Hamilton, Ontario,
                 Canada. aniakata@gmail.com},
  journal     = {Vaccine},
  month       = {May},
  number      = {25},
  pages       = {3778--3789},
  title       = {Anti-vaccine activists, {W}eb 2.0, and the postmodern
                 paradigm--an overview of tactics and tropes used
                 online by the anti-vaccination movement.},
  volume      = {30},
  year        = {2012},
  doi         = {10.1016/j.vaccine.2011.11.112},
  language    = {eng},
  url         = {http://dx.doi.org/10.1016/j.vaccine.2011.11.112}
}

@article{MetcalfCohenLesslerMcAnerneyEtAl2013,
  author      = {Metcalf, C.J.E. and Cohen, C. and Lessler, J. and
                 McAnerney, J.M. and Ntshoe, G.M. and Puren, A. and
                 Klepac, P. and Tatem, A. and Grenfell, B.T. and
                 Bj{\o}rnstad, O.N.},
  institution = {Department of Zoology, Oxford University, Oxford, UK.
                 charlotte.metcalf@zoo.ox.ac.uk},
  journal     = {Journal of the Royal Society, Interface},
  month       = {Jan},
  number      = {78},
  pages       = {20120756},
  title       = {Implications of spatially heterogeneous vaccination
                 coverage for the risk of congenital rubella syndrome
                 in {S}outh {A}frica.},
  volume      = {10},
  year        = {2013},
  doi         = {10.1098/rsif.2012.0756},
  language    = {eng},
  url         = {http://dx.doi.org/10.1098/rsif.2012.0756}
}

@article{SattenspielDietz1995,
  author  = {Sattenspiel, L. and Dietz, K.},
  journal = {Mathematical Biosciences},
  pages   = {71-91},
  title   = {A structured epidemic model incorporating geographic
             mobility among regions},
  volume  = {128},
  year    = {1995}
}

@article{SattenspielHerring2003,
  author  = {Sattenspiel, L. and Herring, D.A.},
  journal = {Bull. Math. Biol.},
  pages   = {1-26},
  title   = {Simulating the effect of quarantine on the spread of
             the 1918-19 flu in central {C}anada},
  volume  = {65},
  year    = {2003}
}

@article{ArinoVdD2003b,
  author    = {Arino, J. and van den Driessche, P.},
  journal   = {Lecture Notes in Control and Information Science},
  pages     = {135--142},
  title     = {The basic reproduction number in a multi-city
               compartmental epidemic model},
  volume    = {294},
  year      = {2003},
  doi       = {10.1007/978-3-540-44928-7_19},
  url       = {ArinoVdD-2003-LNCIS294.pdf}
}

@article{ArinoVdD2003a,
  author   = {Arino, J. and van den Driessche, P.},
  journal  = {Mathematical Population Studies},
  number   = {3},
  pages    = {175--193},
  title    = {A multi-city epidemic model},
  volume   = {10},
  year     = {2003},
  doi      = {10.1080/08898480306720},
  url      = {ArinoVdD-2003-MPS10.correct.pdf}
}

@article{IggidrSalletTsanou2012,
  author    = {Iggidr, A. and Sallet, G. and Tsanou, B.},
  journal   = {Mathematical Population Studies},
  number    = {3},
  pages     = {115-129},
  title     = {Global stability analysis of a metapopulation {SIS}
               epidemic model},
  volume    = {19},
  year      = {2012}
}

@inproceedings{Arino2009,
  author    = {Arino, J.},
  booktitle = {Modeling and Dynamics of Infectious Diseases},
  editor    = {Ma, Z. and Zhou, Y. and Wu, J.},
  pages     = {65--123},
  publisher = {World Scientific},
  series    = {{Series in Contemporary Applied Mathematics}},
  title     = {Diseases in metapopulations},
  volume    = {11},
  year      = {2009},
  url       = {Arino_metapopulations.pdf}
}

@article{ArinoVdD2006a,
  author    = {Arino, J. and van den Driessche, P.},
  journal   = {Fields Institute Communications},
  pages     = {1--12},
  title     = {{Metapopulations epidemic models. A survey}},
  volume    = {48},
  year      = {2006},
  url       = {ArinoVdD-2006-FIC48.pdf}
}

@article{ArinoBajeuxKirkland2019,
  author    = {Arino, J. and Bajeux, N. and Kirkland, S.},
  journal   = {Bulletin of Mathematical Biology},
  number    = {6},
  pages     = {1916--1942},
  title     = {Number of source patches required for population
               persistence in a source-sink metapopulation},
  volume    = {81},
  year      = {2019},
  doi       = {10.1007/s11538-019-00593-1},
  url       = {https://rdcu.be/bpUvv}
}

@book{DiekmannHeesterbeek2000,
  author    = {Diekmann, O. and Heesterbeek, J.A.P.},
  publisher = {Wiley},
  title     = {Mathematical Epidemiology of Infectious Diseases:
               Model Building, Analysis and Interpretation},
  year      = {2000}
}

@article{VdDWatmough2002,
  author  = {van den Driessche, P. and Watmough, J.},
  journal = {Mathematical Biosciences},
  pages   = {29-48},
  title   = {Reproduction numbers and sub-threshold endemic
             equilibria for compartmental models of disease
             transmission},
  volume  = {180},
  year    = {2002}
}

@article{Thieme1993,
  author    = {Thieme, H.R.},
  journal   = {SIAM J. Math. Anal.},
  pages     = {407--435},
  title     = {Persistence under relaxed point-dissipativity (with
               applications to an endemic model)},
  volume    = {24},
  year      = {1993}
}

@article{ArinoMccluskeyVdD2003,
  author    = {Arino, J. and McCluskey, C.C. and
               van den Driessche, P.},
  journal   = {SIAM J Appl Math},
  number    = {1},
  pages     = {260--276},
  title     = {Global results for an epidemic model with vaccination
               that exhibits backward bifurcation},
  volume    = {64},
  year      = {2003},
  doi       = {10.1137/S0036139902413829},
  url       = {ArinoMccluskeyVdD-2003-SIAP64_1.pdf}
}

@article{ArinoDucrotZongo2012,
  author    = {Arino, J. and Ducrot, A. and Zongo, P.},
  journal   = {Journal of Mathematical Biology},
  number    = {3},
  pages     = {423--448},
  title     = {A metapopulation model for malaria with
               transmission-blocking partial immunity in hosts},
  volume    = {64},
  year      = {2012},
  doi       = {10.1007/s00285-011-0418-4},
  url       = {ArinoDucrotZongo-2012-JMB64.pdf}
}

@book{YinZhang1998,
  author    = {Yin, G.G. and Zhang, Q.},
  publisher = {Springer},
  title     = {Continuous-Time Markov Chains and Applications: A
               Singular Perturbation Approach},
  year      = {1998},
  doi       = {10.1007/978-1-4612-0627-9}
}

@book{Kato1995,
  author    = {Kato, T.},
  edition   = {Reprint of the 1980 Second},
  publisher = {Springer-Verlag},
  title     = {Perturbation Theory for Linear Operators},
  year      = {1995}
}

@book{Smith1995,
  author    = {Smith, H.L.},
  title     = {Monotone Dynamical Systems: An Introduction to the Theory of Competitive and Cooperative Systems},
  series    = {Mathematical Surveys and Monographs},
  volume    = {41},
  publisher = {American Mathematical Society},
  address   = {Providence, RI},
  year      = {1995}
}

@book{HornJohnson1990,
  author    = {Horn, R. and Johnson, C.},
  publisher = {Cambridge University Press},
  title     = {Matrix Analysis},
  year      = {1990}
}

@misc{arino_wikipediagatn,
  author       = {Arino, J. and Ciupeanu, A.-S.},
  howpublished = {\url{https://github.com/julien-arino/wikipediaGATN}},
  publisher    = {pyOpenSci},
  title        = {{wikipediaGATN}},
  year         = {2026}
}

@misc{cdc_polio_disease,
  author       = {{Centers for Disease Control and Prevention}},
  howpublished = {\url{https://www.cdc.gov/polio/what-is-polio/index.html}},
  title        = {Polio Disease and Poliovirus}
}

@misc{who_polio_factsheet,
  author       = {{World Health Organization}},
  howpublished = {\url{https://www.who.int/news-room/fact-sheets/detail/poliomyelitis}},
  title        = {Poliomyelitis Fact Sheet}
}

@article{FineCarneiro1999,
  author  = {Fine, Paul E. M. and Carneiro, Ilona A. M.},
  title   = {Transmissibility and Persistence of Oral Polio Vaccine Viruses: Implications for the Global Poliomyelitis Eradication Initiative},
  journal = {American Journal of Epidemiology},
  volume  = {150},
  number  = {10},
  pages   = {1001--1021},
  year    = {1999},
  doi     = {10.1093/oxfordjournals.aje.a009924}
}

@article{DuintjerTebbens2013,
  author  = {Duintjer Tebbens, Radboud J. and Pallansch, Mark A. and Kalkowska, Dominika A. and Wassilak, Steven G. F. and Cochi, Stephen L. and Thompson, Kimberly M.},
  title   = {Characterizing Poliovirus Transmission and Evolution: Insights from Modeling Experiences with Wild and Vaccine-Related Polioviruses},
  journal = {Risk Analysis},
  volume  = {33},
  number  = {4},
  pages   = {703--749},
  year    = {2013},
  doi     = {10.1111/risa.12044}
}

@misc{phac_measles_2025,
  author       = {{Public Health Agency of Canada}},
  title        = {Statement from the {Public Health Agency of Canada} on {Canada's} Measles Elimination Status},
  howpublished = {\url{https://www.canada.ca/en/public-health/news/2025/11/statement-from-the-public-health-agency-of-canada-on-canadas-measles-elimination-status.html}},
  month        = nov,
  year         = {2025}
}

\end{document}